\documentclass[twocolumn,longauthor]{aastex63}

\newcommand{\gaia}{{\it Gaia} }

\newcolumntype{?}{!{\vrule width 1pt}}

\usepackage{booktabs, tabularx, makecell, multirow, xspace}
\newcolumntype{C}[1]{>{\centering\let\newline\\\arraybackslash\hspace{0pt}}m{#1}}

\DeclareUnicodeCharacter{2212}{-}

\shorttitle{}
\shortauthors{Dropulic et al.}

\graphicspath{{./}{figures/}}
\usepackage{url}
\usepackage[normalem]{ulem}
\usepackage{amsmath}
\usepackage{comment}
\usepackage{svg}
\usepackage{natbib}
\usepackage{fontawesome}

\newcommand{\zenodo}{\href{https://zenodo.org/record/6558083}{\faCloudDownload}\xspace}

\begin{document}

\title{\vspace{-40pt} Revealing the Milky Way’s Most Recent Major Merger \\ with a \textbf{\textit{Gaia}} EDR3 Catalog of Machine-Learned Line-of-Sight Velocities}

\correspondingauthor{Adriana Dropulic}
\email{dropulic@princeton.edu}

\author[0000-0002-7352-6252]{Adriana Dropulic}
\affiliation{Department of Physics, Princeton University, Princeton, NJ 08544, USA }

\author[0000-0003-2486-0681]{Hongwan Liu}
\affiliation{Department of Physics, Princeton University, Princeton, NJ 08544, USA }
\affiliation{Center for Cosmology \& Particle Physics, Department of Physics, New York University, New York, NY 10003, USA}

\author[0000-0002-0376-6461]{Bryan Ostdiek}
\affiliation{Department of Physics, Harvard University, Cambridge, MA 02138, USA}
\affiliation{The NSF AI Institute for Artificial Intelligence and Fundamental Interactions}

\author[0000-0002-8495-8659]{Mariangela Lisanti}
\affiliation{Department of Physics, Princeton University, Princeton, NJ 08544, USA }
\affiliation{Center for Computational Astrophysics, Flatiron Institute, 162 Fifth Ave, New York, NY 10010, USA}

\begin{abstract}
Machine learning can play a powerful role in inferring missing line-of-sight velocities from astrometry in surveys such as \emph{Gaia}. In this paper, we apply a neural network to \gaia Early Data Release~3~(EDR3) and obtain line-of-sight velocities and associated uncertainties for $\sim 92$ million stars.  The network, which takes as input a star's parallax, angular coordinates, and proper motions, is trained and validated on $\sim 6.4$ million stars in \gaia with complete phase-space information. The network's uncertainty on its velocity prediction is a key aspect of its design; by properly convolving these uncertainties with the inferred velocities, we obtain accurate stellar kinematic distributions.  As a first science application, we use the new network-completed catalog to identify candidate stars that belong to the Milky Way's most recent major merger, \textit{Gaia}-Sausage-Enceladus~(GSE). We present the kinematic, energy, angular momentum, and spatial distributions of the $\sim 450,000$ GSE candidates in this sample, and also study the chemical abundances of those with cross matches to GALAH and APOGEE.  The network's predictive power will only continue to improve with future \emph{Gaia} data releases as the training set of stars with complete phase-space information grows. This work provides a first demonstration of how to use machine learning to exploit high-dimensional correlations on data to infer line-of-sight velocities, and offers a template for how to train, validate and apply such a neural network when complete observational data is not available. \zenodo
\vspace{30pt}
\end{abstract}

\section{Introduction} \label{sec:intro}

The Milky Way galaxy formed through a process of hierarchical structure formation, capturing and dismantling smaller galaxies with its strong gravitational pull~\citep{white_galaxy_1991}.  Stars accreted from a disrupted galaxy may retain characteristic chemical and phase-space features, which can be harnessed to infer properties about the parent galaxy~\citep{Helmi_2003, Bullock_2005, Robertson_2005, Font_2006, DeLucia_2008}.
This effort in galactic archaeology is greatly aided by the availability of complete velocity information for the Milky Way's stars.  In this work, we use a neural network to infer missing line-of-sight velocities from available 5D astrometry in \gaia Early Data Release~3~(EDR3)~\citep{gaia_collaboration_gaia_2021}, resulting in $\sim$ 92 million more stars with  estimated 6D phase-space coordinates.  As a first science application, we use the catalog of machine-learned line-of-sight velocities to study the properties of one of the most significant mergers in the Milky Way's history, \textit{Gaia}-Sausage-Enceladus~(GSE)~\citep{belokurov_co-formation_2018,helmi_merger_2018}.

 The \gaia satellite has ushered in a new age in astrometry, with the goal of providing precise positions and velocities for an unprecedented number of stars in the Milky Way~\citep{2016A&A...595A...1G,gaia_collaboration_gaia_2018}.   However, most stars in \emph{Gaia} EDR3 have only 5D astrometry available (two angular coordinates, two proper motions, and parallax).  Currently, less than 1\% of the stars in the \gaia catalog have a measured line-of-sight velocity, but this will continue improving with the addition of $\sim 30$~million more line-of-sight velocities in the next data release~\citep{gaia_collaboration_gaia_2021}.  Spectroscopic surveys play an important role in complementing the \gaia data with high-precision line-of-sight velocities and chemical abundances.  However, these surveys typically have limited sky coverage and only a small fraction of stars are cross-matched with \textit{Gaia}.  As we will show, using a neural network  
 to fill in missing phase-space information in astrometric data
is a powerful alternative when complete observational data is not available.  

\cite{dropulic_machine_2021} (hereafter referred to as D21)
 developed a deep neural network design that accepted 5D astrometric coordinates (angular coordinates, proper motions, and parallax) and output the missing line-of-sight velocity, $v_{\rm los}$, along with an uncertainty on the predicted velocity, $\sigma_{\rm los}$.  
 D21 trained, tested, and validated the neural network on mock Milky Way stellar data.  The use of simulations allowed for clear metrics of success to guide the network development because truth information was available.  
To assess the network's predictive power, D21 compared the true line-of-sight velocities to the predicted ones.  They found that 5D astrometry alone is insufficient to provide a reliable prediction of $v_{\rm los}$ for every individual star.  However, incorporating the learned uncertainty, $\sigma_{\rm los}$, allowed them to obtain accurate \textit{distributions} for the full population of stars, as well as the correlations between different velocity components in the Galactocentric frame.  Thus, the uncertainty output of the network was a crucial feature of its design.  Moreover, D21 showed that the neural network was successful in accurately reconstructing spatially diffuse kinematic substructure, like the GSE, that comprises a small fraction of the overall data. This general idea of inferring missing line-of-sight velocities based on information acquired from a smaller subset of stars with full 6D information has recently been applied to stars in the Kepler field~\citep{2022arXiv220508901A}, albeit using a Bayesian inference approach. 

Motivated by the results of D21, we train a neural network on $\sim 6.4$~million \gaia EDR3 stars with line-of-sight velocities and then use it to predict the line-of-sight velocities, and associated uncertainties, of an additional $\sim 92$ million stars. The network is validated by comparing its predictions to the measured values for the subset of data with full phase-space information.\footnote{Note that the training, testing, and validation  of the network are all performed on \gaia EDR3 data in this work.  While mock data catalogs were used by D21 to develop the network architecture, they are not used at any stage of the analysis here.}   This work provides a concrete demonstration of how machine learning can be used to fill in stellar phase-space in future \gaia data releases.  Indeed, as the subset of measured \gaia  line-of-sight velocities continues to grow in the years ahead, so too will the training set, and the prediction accuracy of the network should only continue to improve.

As a first application, we use the catalog of stars with machine-learned line-of-sight velocities to identify candidates of the GSE merger.  To date, most studies of GSE have been restricted to a small fraction of stars from \gaia with 6D phase-space or that have been cross-matched with spectroscopic surveys. Still, much has been learned about the GSE merger~\citep{helmi_streams_2020}.  GSE tidal debris is thought to comprise a significant part of the Galactic stellar halo.  It can potentially explain the observed break in the halo density and anisotropy profiles at Galactocentric radii of $\sim 20$--30~kpc; the anisotropic component is thought to be the tidal debris of GSE, which is mixed together with an isotropic, metal-poor component of the Galactic halo~\citep{deason_apocenter_2018,Haywood_2018, lancaster_halos_2019, bird_anisotropy_2019, iorio_chemo-kinematics_2021}. Close to the Solar position, estimates suggest that the GSE debris can comprise more than 50\%, and possibly up to 80\%, of the stellar halo~\citep{necib_under_2019, lancaster_halos_2019,naidu_evidence_2020,iorio_chemo-kinematics_2021}.  
GSE stars are on highly eccentric, radial orbits, and the original satellite likely had a total stellar mass of $M_{*} \sim 10^{8-9} M_{\odot}$ at infall~\citep{koppelman_one_2018, 10.1093/mnrasl/slz070, 2019MNRAS.488.1235M, mackereth_origin_2019, 10.1093/mnras/staa047, koppelman_massive_2020,yuan_low-mass_2020, 2020MNRAS.498.2472K,iorio_chemo-kinematics_2021,carollo_nature_2021,limberg_targeting_2021}.  Moreover, the metallicity distribution of GSE stars is estimated to peak near $\text{[Fe/H]}\sim -1.4$ to $-1.2$, consistent with the picture that the collision between GSE and the Milky Way occurred $\sim8$ to 10~Gyr ago~\citep{Myeong_2018,2019NatAs...3..932G,2019A&A...630L...4M,mackereth_origin_2019,necib_under_2019,2019A&A...632A...4D,10.1093/mnrasl/slz070,bonaca_timing_2020,naidu_evidence_2020,das_ages_2020,10.1093/mnras/staa047,2020MNRAS.493..847F,10.1093/mnras/staa1888,feuillet_selecting_2021,Hasselquist_2021,gudin_r-process_2021,2021arXiv211011465D,2021NatAs...5..640M,2022MNRAS.510.2407B,2022arXiv220404233H}. 

Our network-completed catalog allows us to identify nearly twenty-fold more GSE candidate stars than possible from the 6D \gaia dataset alone---nearly $450,000$ stars in total.  We then use these stars to infer the metallicity distribution function for stars that are cross-matched with spectroscopic surveys and the spatial distribution of GSE for all stars. In both cases, we find excellent agreement between the true and predicted distributions for stars with full 6D information.  For stars with only 5D information, the predicted metallicity distribution as a function of $\text{[Fe/H]}$ and $\text{[Mg/Fe]}$ for candidate GSE stars are consistent with both the 6D results and existing studies in the literature. The spatial distribution of GSE candidate stars with only 5D information exhibits hints of anisotropy in Galactocentric $y_\mathrm{gal}-z_\mathrm{gal}$ coordinates, as compared to the mostly isotropic distribution recovered in the 6D dataset.

This paper is organized as follows. In Sec.~\ref{sec:edr3}, we review the neural network developed in D21, overview the necessary changes for its application to \gaia EDR3 data, and discuss validation of the network output. In Sec.~\ref{sec:mlcat}, we introduce the network-completed \gaia EDR3 catalog. In Sec.~\ref{sec:enc}, we present our GSE selection technique and discuss the dynamics, metallicity, and spatial distribution of the candidate stars.  We conclude in Sec.~\ref{sec:conclude}. 

The catalog of machine-learned line-of-sight velocities and uncertainties for \gaia EDR3 together with their source IDs has been made publicly available. \zenodo

\section{Network Testing and Validation} \label{sec:edr3}

\begin{table*}[t!]
\centering
\footnotesize
\renewcommand{\arraystretch}{2}
\noindent
\begin{tabular}{ C{0.5 cm} | C{5.5 cm} | C{2 cm}   C{2 cm}  C{2 cm}}
\Xhline{3\arrayrulewidth}
 & Selection Cuts & EDR3 RV \newline Catalog & Test-Set Catalog &ML-RV Catalog \\
  \hline
  0 & $\varpi > 0$, $\delta\varpi/\varpi<0.2$, $G\in [6,17]$, $\text{ruwe} < 2$ & \textbf{6,444,871} & \textbf{6,444,871} & 92,371,404\\ 
 1 & [0] + $G\in [12,17]$ & 4,332,657 & 4,332,657 & \textbf{91,840,346}\\
 2 & [1] + $(e > 0.8)$  &  24,219 & $\sim18,000$ & $\sim450,000$\\  
 3 & [2] + GALAH  & 244 & $\sim200$ & $\sim400$\\  
 4 & [2] + APOGEE & 624 & $\sim500$ & $\sim800$\\ 
 \Xhline{3\arrayrulewidth}
\end{tabular}
\vspace{5mm}

\caption{Number of stars in various subsets of \emph{Gaia} EDR3 used in this work.  The subsets include the Radial Velocity~(RV) Catalog, the Test-Set Catalog, and the Machine Learned-Radial Velocity~(ML-RV) Catalog, with the total number of stars in each catalog indicated in bold.  The first is simply the subset of \emph{Gaia} EDR3 data with measured line-of-sight velocities~\citep{katz_gaia_2019}, which also satisfy the base selection requirements in Row~[0] of the table in addition to having $v_\text{los} \in [-550, 550]~\text{km/s}$.  The Test-Set Catalog is similar to the RV~Catalog, except with line-of-sight velocities and associated uncertainties inferred by the neural network.  The ML-RV Catalog is the subset of \emph{Gaia} EDR3 that also satisfies the selection requirements in Row~[0] in addition to $G \in [12,17]$ (Row [1]), but with no measured line-of-sight velocities.  Each star in the ML-RV Catalog has a line-of-sight velocity and an associated uncertainty predicted by the neural network.  Additional selection cuts include [2]~cutting on orbital eccentricity, $e$, [3]~cross-matching with GALAH\texttt{+}~DR3~\citep{2020arXiv201102505B}, and [4]~cross-matching with APOGEE~DR17~\citep{2017AJ....154...94M}.  Note that $e$ corresponds to the true eccentricity for the RV~Catalog and the predicted value for the Test-Set and ML-RV Catalogs.  The number of stars with $e > 0.8$ for the Test-Set and ML-RV Catalogs are approximate, since this quantity changes between Monte Carlo draws during error sampling.} 
\label{table:1}
\end{table*}

The \emph{Gaia}~catalog contains full phase-space information for several million EDR3 stars~\citep{katz_gaia_2019}.  These stars serve as a critical tool for training, validating, and testing the neural network architecture from D21 in a purely data-driven fashion.  Moving forward, we will work with this subset of  \emph{Gaia} stars, further selecting those with parallax $\varpi > 0$ and fractional error $\delta\varpi/\varpi<0.2$, renormalized unit weight error $\text{ruwe} < 2$, as well as \emph{Gaia}-measured line-of-sight velocity $v_{\rm los} \in [-550,550]$~km/s. Additionally, we restrict to stars with $G$-band magnitude $G \in [6, 17]$ for training.\footnote{Note that when studying GSE candidates later on, we will further restrict to stars with $G \in [12, 17]$ where the EDR3 catalog is essentially spatially complete~\citep{2021A&A...649A...5F}.}  We also correct for the zero-point parallax offset in the \gaia data, which results at least in part from a degeneracy between the parallax and basic-angle variation on the \gaia satellite. Specifically, we use the parallax zeropoint for EDR3 published by the Gaia Collaboration~\citep{2021A&A...649A...4L}, which is a function of $G$, the ecliptic latitude, and the effective wavenumber used in the astrometric solution; the  zeropoint ranges from $-0.08$ to $0.01$. We refer to this subset of \emph{Gaia}, which comprises more than 6.4~million stars, as the \emph{Radial Velocity~(RV)~Catalog}.  A summary of the number of stars in the RV~Catalog as a function of various selection cuts is provided in Tab.~\ref{table:1}.

We train, validate, and test the network on the RV~Catalog and present the results in this section.  Sec.~\ref{sec:architecture} introduces the network architecture used in this study in detail and discusses the error-sampling procedure used in presenting results throughout the paper.  Sec.~\ref{sec:validate} describes how well the network does in predicting the kinematic and spatial distributions of high-eccentricity stars in the RV~Catalog, which are potential GSE candidates.  These results are critical for building confidence in the neural network behavior on data, so that we can move forward and study its results on the \emph{Gaia} data with only 5D astrometry available. 

\subsection{Network Architecture and Error Sampling}\label{sec:architecture}

\begin{figure*}[t]
\centering
\includegraphics[width=\textwidth]{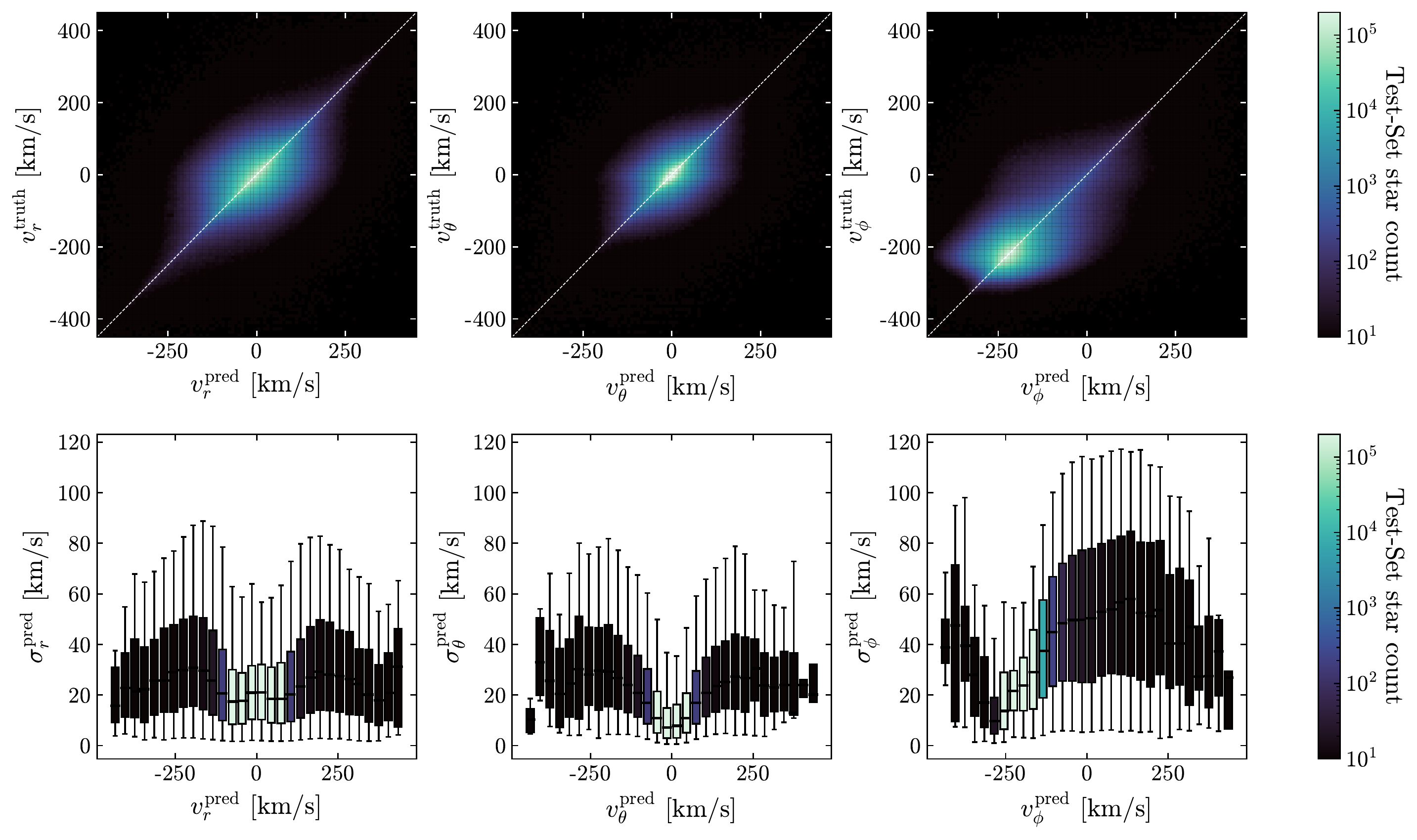}
\caption{\emph{(Top)} Comparison of the inferred velocities, $v^\text{pred}$, in the Test-Set Catalog with their corresponding truth values, $v^\text{truth}$, from the RV~Catalog.  While the network infers the line-of-sight velocities, we present the results here in Galactocentric spherical coordinates.  The dotted white diagonal line indicates exact agreement between the truth and inferred values.  \emph{(Bottom)} The network-predicted uncertainty as a function of velocity in each coordinate direction.  In each bin, the box denotes 50\% containment about the median uncertainty, while the whiskers indicate 5 to 95\% containment.  In this and all following figures, the Test-Set star counts for the background heatmaps are properly error-sampled, as described in the text.}
\label{fig:recap1}
\end{figure*}

The neural network developed in D21 consists of two halves that are structured identically except for the last layer. Each half of the network consists of six layers: the input, four hidden layers, and the output. The input consists of five quantities per star: Galactic longitude~($\ell$), Galactic latitude~($b$), proper motion in right ascension~($\mu_\alpha$), proper motion in declination~($\mu_\delta$), and parallax~($\varpi$). The hidden layers comprise 30 nodes each and use a hyperbolic tangent activation function. The output layer from one half of the network, which we call the “velocity predictor,” consists of a single node with linear activation to obtain a continuous value, the line-of-sight velocity. The output layer from the other half of the network, which we call the
“uncertainty predictor,” uses a ReLU activation function to constrain the uncertainty on the predicted line-of-sight velocity to positive values.

\begin{figure*}
    \centering
    \begin{minipage}{0.47\textwidth}
        \centering
        \includegraphics[width=0.8\textwidth]{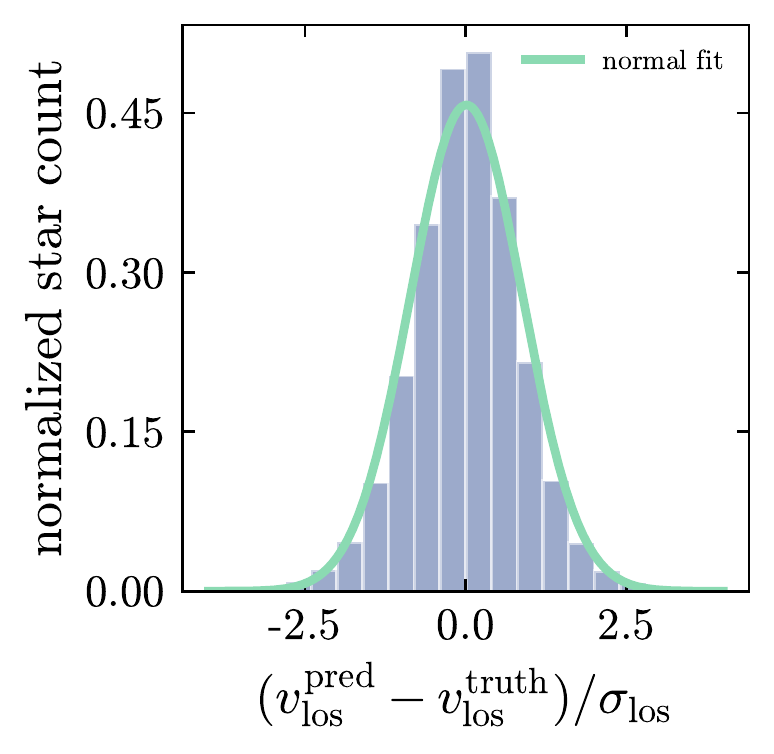} 
        \caption{The distribution of the residual between the network-predicted line-of-sight velocity $v_{\rm los}^{\rm pred}$ and the true value $v_{\rm los}^{\rm truth}$, divided by the network-predicted uncertainty $\sigma_{\rm los}$.  The solid green line shows a best-fit normal distribution with mean $\mu = 0.01$ and standard deviation $\sigma = 0.9$. This justifies interpreting $\sigma_{\rm los}$ as a Gaussian error bar.}
        \label{fig:gauss_uncert}
    \end{minipage}\hfill
    \begin{minipage}{0.47\textwidth}
        \centering
        \includegraphics[width=\textwidth]{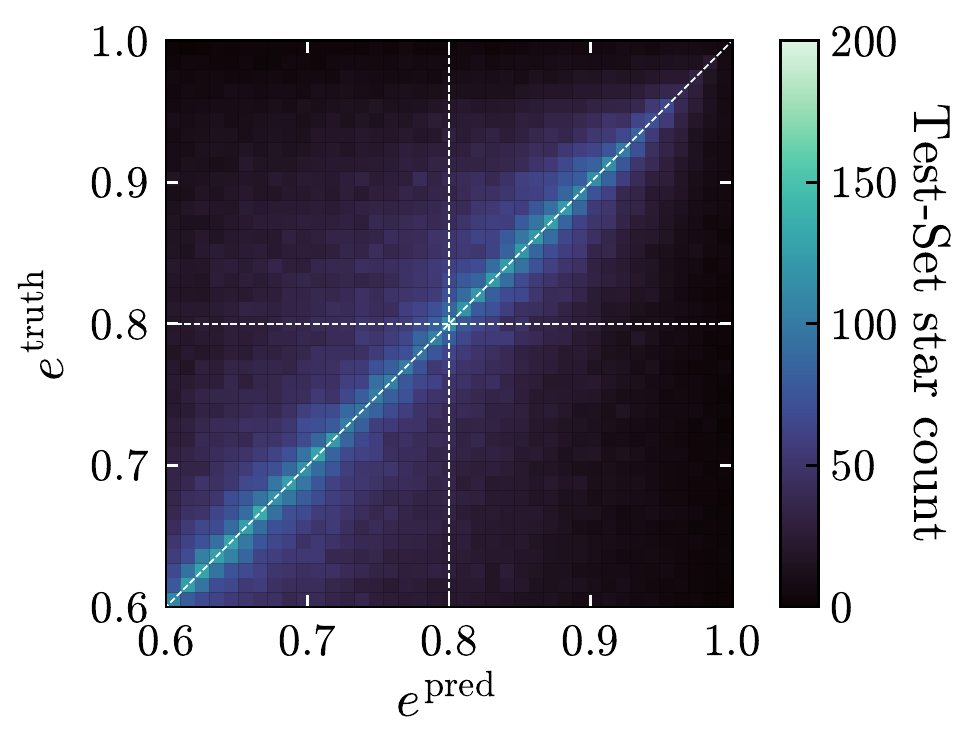} 
        \caption{The network-predicted eccentricities for stars in the error-sampled Test-Set Catalog compared to their true values from the RV~Catalog.  Overall, the correspondence between truth and predicted values is excellent, especially at the highest eccentricities.  Stars that fall in the bottom right~(top left) quadrant would be mistakenly labeled~(not labeled) as GSE candidates in the Test-Set Catalog, assuming a selection cut of $e>0.8$.}
        \label{fig:ecc}
    \end{minipage}
\end{figure*}

To improve the predictive power of the network on \emph{Gaia} EDR3, we make several modifications to the foundational neural network of D21:
\begin{itemize}
\item We provide the network with a continuous representation of Galactic longitude, $\ell$, as opposed to the discontinuous representation $\ell \in [0^{\circ},360^{\circ}]$ used in D21.   Specifically, we use $\ell \rightarrow [\cos(\ell), \sin(\ell)]^{T}$.  This change in representation ameliorates a discontinuity in the neural network's outputs at $\ell = 0^{\circ}$, which was not present in the outputs when the network was trained on the mock catalog.  How the topology of a dataset affects the behavior of the neural network is a novel field of research---see, for example,~\cite{batson_topological_2021}. 
\item We account for measurement errors on the line-of-sight velocity in the \gaia EDR3 RV~Catalog by Monte-Carlo~(MC) sampling each star over this uncertainty during training.  This ensures that the network's predicted uncertainty, $\sigma_{\rm los}$, encompasses the \gaia measurement error. 
\end{itemize}
In all other aspects, the neural network used in this work is the same as that developed in D21.

The RV~Catalog is divided into five random subsets. For each round of training, we choose three different subsets to be the training, validation, and test sets, respectively.\footnote{The training, validating, and testing of the network occur in the same manner as described in D21.} For each round of training, we predict line-of-sight velocity values and uncertainties for all stars in one of the subsets (where truth-level $v_{\rm los}$ is withheld). After all rounds of training are complete, every star in the RV~Catalog has a network-predicted $v_{\rm los}$ and $\sigma_{\rm los}$.  We refer to this as the \emph{Test-Set Catalog}, as it will allow us to compare network predictions for the data to their true values.  By construction, it has the same number of stars as the RV~Catalog.  Properties of the Test-Set Catalog are also summarized in Tab.~\ref{table:1}.

To fully exploit the network prediction of both $v_\mathrm{los}$ and $\sigma_\mathrm{los}$, we perform a procedure we term \emph{error-sampling} to present results throughout this paper. For each star, we take the predicted $v_\mathrm{los}$ and $\sigma_\mathrm{los}$ as the mean and standard deviation of a Gaussian distribution for the line-of-sight velocity and draw five MC trials from this distribution. Any kinematics cut that we impose is performed on each MC sample separately. The final predicted distribution is then taken to be the mean of all the MC trials and corresponds to what is presented in all of the plots to follow. Figure colorbars that are labeled as ``Test-Set star count'' or ``ML-RV star count'' represent error-sampled values.  Colorbars labeled as ``RV star count'' represent truth-level counts.

Figure~\ref{fig:recap1} demonstrates how well the network does in filling in missing phase-space information by comparing predicted velocities in the Test-Set Catalog to their corresponding truth values in the RV~Catalog.  While the network predicts $v_{\rm los}$, we have transformed the results into Galactocentric spherical velocities ($v_{r}, v_{\theta} ,v_{\phi}$), following the same procedure described in Appendix B of D21.
The top row of Fig.~\ref{fig:recap1} compares the predicted velocities to their truth values in the RV~Catalog. Perfect agreement between truth and predicted velocities occur along the diagonal, with the prediction being an underestimate~(overestimate) toward the left~(right) of the plot. The bottom row shows how the uncertainties depend on the predicted velocities. The predicted uncertainty on the Galactocentric velocities is shown as a function of the predicted Galactocentric velocity for all stars in the Test-Set~Catalog. For each velocity bin, the box denotes the 50\% containment about the median uncertainty, while
the whiskers denote the 5\% and 95\% containment. The network uncertainties are lowest for values of ($v_r, v_\theta, v_\phi$) corresponding to disk stars, since they dominate the catalog in sheer number. Because non-disk stars form a comparatively smaller fraction of the data, the network uncertainties are typically larger.  However, these uncertainties in the network output are taken into account through the \emph{error-sampling} procedure.

Figure~\ref{fig:gauss_uncert} explicitly demonstrates that the network's predicted uncertainties, $\sigma_{\rm los}$, can be interpreted as Gaussian.  Here, we histogram the residuals of the predicted line-of-sight velocities relative to their corresponding truth values, divided by the predicted uncertainty.  The distribution is well-fit by a Gaussian with mean $\mu=0.01$ and standard deviation $\sigma = 0.9$. We conclude that approximately 68\%~(95\%) of stars have true line-of-sight velocity $v_\mathrm{los}^\mathrm{truth}$ within $1\sigma_\mathrm{los}$~($2\sigma_\mathrm{los})$ of the network's predicted central value $v_{\rm los}^{\rm pred}$, and that $\sigma_\mathrm{los}$ can be safely interpreted as a Gaussian error bar. 

\begin{figure*}[t]
\centering
\includegraphics[width=\textwidth]{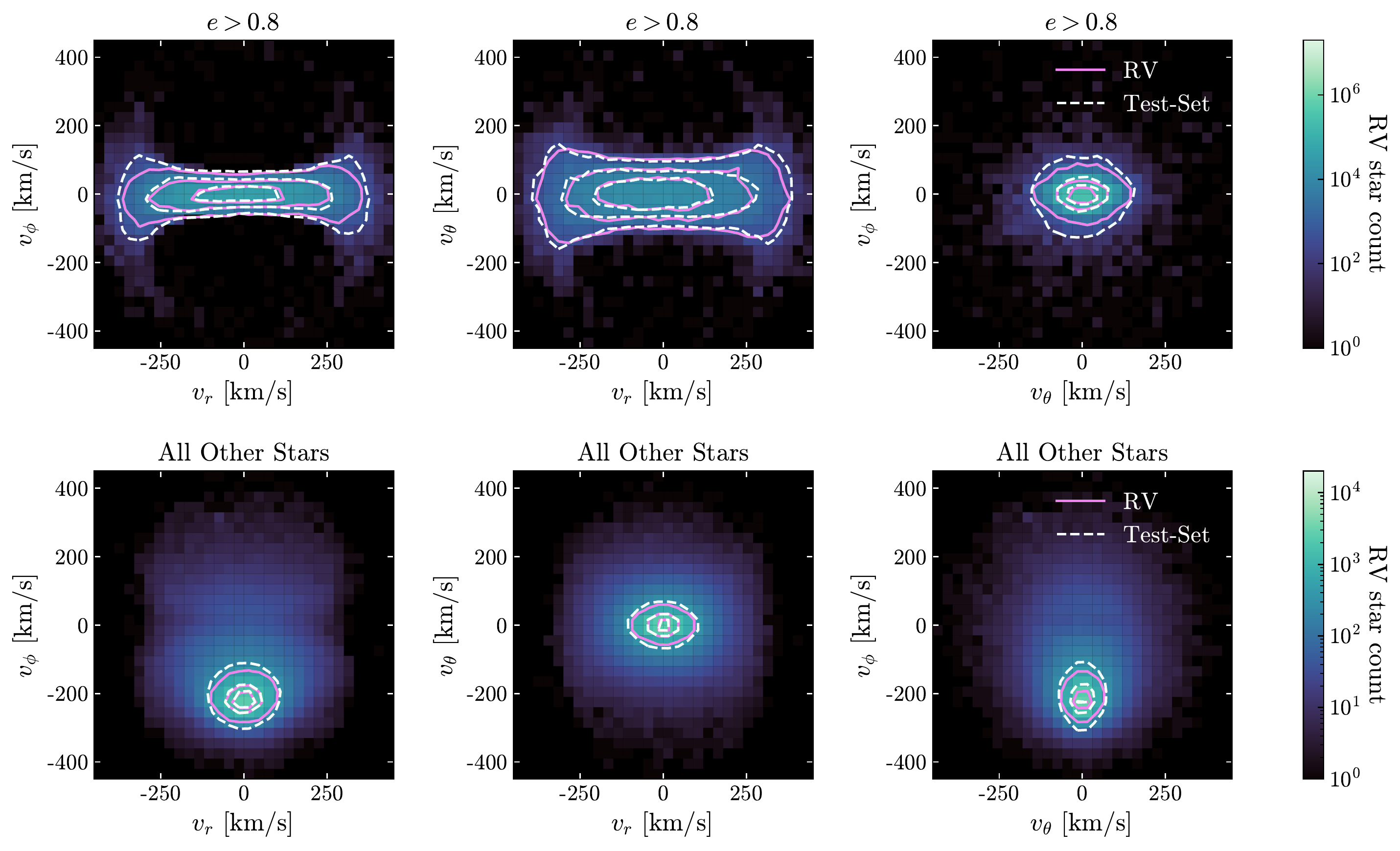}
\caption{Distributions of Galactocentric velocities for the high eccentricity sample \emph{(top)} and all other stars \emph{(bottom)}. The background heatmaps correspond to truth-level distributions from the RV~Catalog.  The corresponding 30\%, 60\%, and 90\% containment contours are shown by the solid violet lines.  For comparison, we also show the error-sampled containment regions for stars in the Test-Set~Catalog in dashed white.  The correspondence between the true and machine-learned distributions is excellent. Note that $e$ refers to the true~(predicted) eccentricities for stars in the RV~(Test-Set)~Catalog.}
\label{fig:vel5D}
\end{figure*} 

\subsection{High-Eccentricity Stars in the Test-Set Catalog}
\label{sec:validate}

GSE stars are typically identified by some combination of the following properties: high velocity, retrograde motion, low metallicity, and highly eccentric orbits. Our goal is to identify candidate GSE stars from the subset of \emph{Gaia} data with only 5D astrometry available.  To do so, we select stars on high-eccentricity orbits, as determined using the network-inferred line-of-sight velocities.  To this end, the network must be able to identify the low- and high-eccentricity stellar population accurately.  In this subsection, we validate the performance of the network by comparing the low- and high-eccentricity populations in the RV~Catalog and the Test-Set Catalog. We emphasize that the network does not require any knowledge of the Galactic potential to make the predictions that we detail below. 

\begin{figure*}
   \centering
\begin{tabular}{C{\textwidth}}
\includegraphics[width=\textwidth]{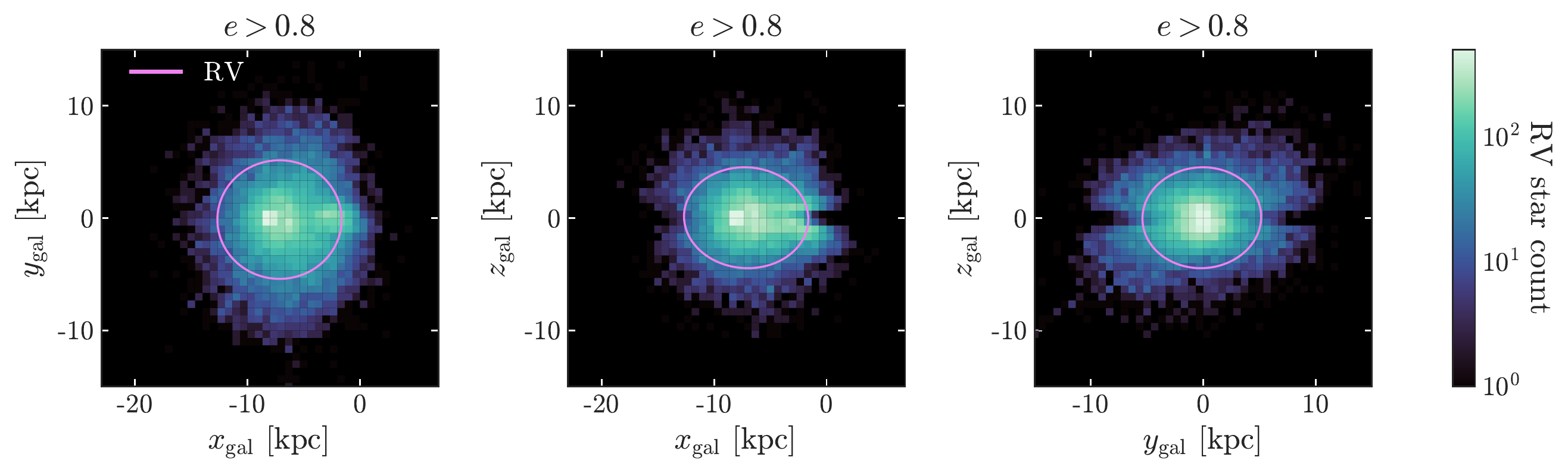}\\
\includegraphics[width=\textwidth]{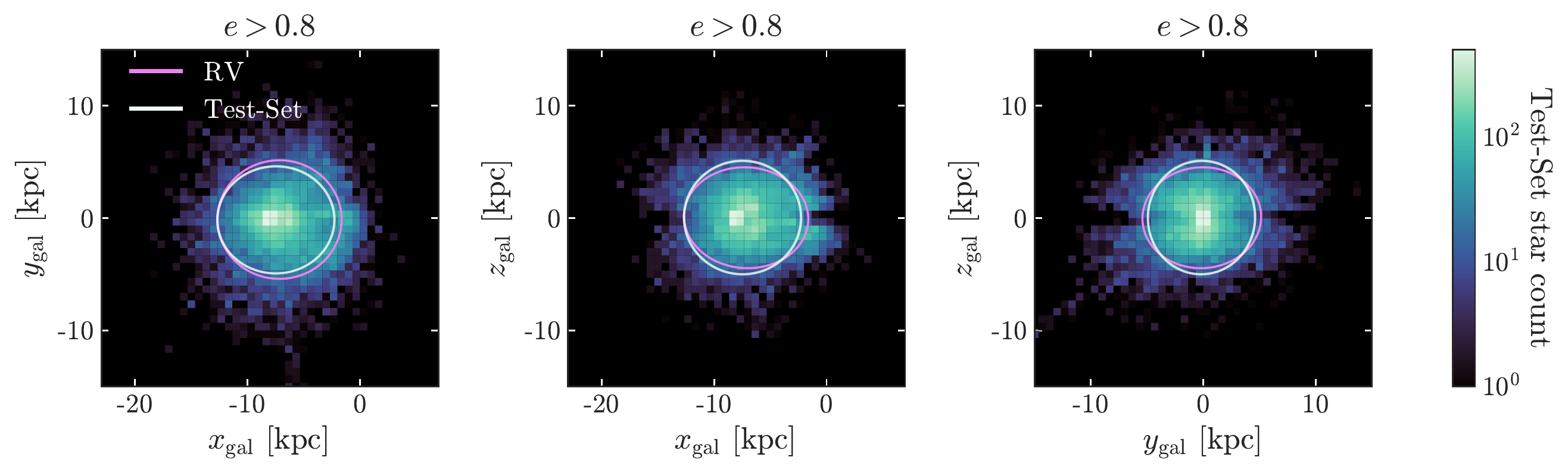}\\
\end{tabular}
    \caption{Heatmaps corresponding to the Galactocentric spatial distributions for the high-eccentricity stars in the RV~Catalog~\emph{(top)} and Test-Set~Catalog~\emph{(bottom)}. In this work, we use a right-handed coordinate system where the Sun is located at negative $x_{\rm gal}$.  The 2D covariance ellipses for all stars within 5~kpc of the Sun are indicated by the violet~(white) lines for the RV~(Test-Set)~Catalog.  There are five essentially indistinguishable lines shown for the fits to the Test-Set~Catalog; each corresponds to a different sample over the network uncertainty.  A summary of the tilt angles and axis ratios of the covariance ellipses are provided in Tab.~\ref{table:2}.}
\label{fig:spatial6D}
\end{figure*}

Moving forward, we will use a high-eccentricity cut of $e>0.8$ to identify GSE candidates.  This selection criterion is similar to that used by~\cite{naidu_evidence_2020}, which required that $e > 0.7$ and also explicitly excluded other pre-defined substructures.  We place a  tighter restriction on eccentricity to decrease contamination from stars with low eccentricities. We find that only $0.15\%$~($0.25\%$) of the Test-Set Catalog stars have $e^{\rm pred} > 0.8~(0.7)$ when their corresponding truth values are less than 0.8~(0.7). It is important to note that, without chemical abundances, it is difficult for us to remove possible contamination from other substructures as done by~\cite{naidu_evidence_2020}.    

To determine stellar eccentricities, we use the \texttt{gala}  package~\citep{m_price-whelan_gala_2017}, assuming the standard Milky Way potential given in~\cite{Bovy_2015}.\footnote{We note that the derived eccentricities may differ if a triaxial halo is used for the orbit integration, as discussed in~\cite{2022arXiv220207662H}.}  The notation $e^{\rm pred}$ refers to eccentricities calculated from the Test-Set Catalog and $e^{\rm truth}$ to eccentricities calculated directly from truth velocities in the RV~Catalog.  The correlation coefficient between $e^{\rm pred}$ using $v_\mathrm{los}^\mathrm{pred}$ and $e^{\rm truth}$ for all stars is 0.82, indicating excellent agreement.  Figure~\ref{fig:ecc} shows the  correlation between $e^{\rm pred}$ and $e^{\rm truth}$ for high-eccentricity stars, including our error-sampling procedure, in the relevant parameter space for the GSE selection. Points along the diagonal indicate a completely accurate predicted eccentricity, while points above (below) the diagonal indicate an underestimated (overestimated) eccentricity. There are notably fewer stars in the bottom right-hand quadrant of the plot; this is important because we will need to cut on $e^{\rm pred}$ to obtain GSE candidates in the full \gaia EDR3 catalog, and this tells us that stars which are assigned higher values of $e^{\rm pred}$ generally have as high or higher $e^{\rm truth}$ values. Of the $\sim 60,000$ stars shown in Fig.~\ref{fig:ecc}, $\sim 28\%$ are in the upper-right quadrant, $\sim 17\%$ are in the upper-left quadrant, and only $ \sim 12\%$ are in the bottom-right quadrant. Of the $ \sim 12\%$ of stars in the bottom-right quadrant of 
Fig.~\ref{fig:ecc} (representing the contamination from low-eccentricity stars in our sample), $ \sim 65\%$ have $e^{\rm truth} > 0.7$, which would still qualify as high-eccentricity per the cut made by~\cite{naidu_evidence_2020}.

Figure~\ref{fig:vel5D} explores the extent to which the network captures correlations among the velocity components for the low and high-eccentricity samples. The background heatmaps correspond to the distribution of stars in the RV~Catalog. Here we begin to use our spatially-complete Catalog; there are 4,308,438 stars with $e^\text{truth} < 0.8$ and 24,219 stars with $e^\text{truth} > 0.8$ (Row~[2] of Tab.~\ref{table:1}).  As expected, the low-eccentricity sample of stars is disk-like with azimuthal velocities strongly peaked at $v_\phi \sim -200$~km/s.  The high-eccentricity sample has the characteristic ``sausage''-like shape in the radial velocity distribution, which is expected for GSE stars~\citep{belokurov_co-formation_2018}. The $v_\phi$ distribution for the low-eccentricity stars is not smooth across $v_\phi \sim 0$~km/s because many of these stars are on highly radial orbits (\emph{i.e.}, have large $v_r$) and thus appear in the top panel instead of the bottom panel of the figure. 

The violet contours in Fig.~\ref{fig:vel5D} correspond to 30\%, 60\% and 90\% containment intervals for stars in the RV~Catalog.  The contours for the corresponding error-sampled distributions of the Test-Set~Catalog are shown in dashed white. Because of our conservative error-sampling procedure, $\sim 3,000$ stars are excluded out of the typical sample size of 18,000 stars in each of the MC'd sets with $e^\text{pred} > 0.8$; these stars have eccentricities that fall below the cut after the MC sampling.
The error-sampled Test-Set Catalog is an excellent approximation of the RV~Catalog for both the high-eccentricity and low-eccentricity samples at the 30\% and 60\% containment intervals.  The network is marginally under-confident at the 90\% level, likely because stars on the distribution tails typically have higher network uncertainties. We stress that the correspondence between the predicted and true velocity distributions is highly nontrivial.  The network has no direct access to information about the Galactic potential, which is essential in predicting its eccentricity; nevertheless, the network's ability to predict $v_\mathrm{los}$ and $\sigma_\mathrm{los}$ is sufficiently good to accurately predict the distributions for these categories. 

\begin{table}[t!]
\footnotesize
\renewcommand{\arraystretch}{1.75}
\begin{center}
\begin{tabular}{C{1.4 cm} | C{2.1 cm} | C{1.8 cm} |  C{1.8 cm}}
  \Xhline{3\arrayrulewidth}
  & RV \newline Catalog & Test-Set Catalog &ML-RV Catalog \\
  \hline
   Tilt Angle & $\sim$ isotropic & $\sim$ isotropic & $59^\circ$ \\ 
 Axis Ratio  & 10.2 : 10.2 : 9.0 & 9.6 : 9.6 : 9.0 & 9.0 : 7.8 : 7.2\\
  \Xhline{3\arrayrulewidth}
\end{tabular}
\end{center}
\caption{Tilt angles and axis ratios for the covariance ellipses of the spatial distributions of all stars within 5~kpc of the Sun in the RV, Test-Set, and ML-RV~Catalog. The covariance ellipses are intended to quantitatively compare the spatial distribution of GSE candidates between the different catalogs; they apply to the local, not global, halo. However, the tilt in the $y_{\rm gal} - z_{\rm gal}$ plane may be an unbiased indicator of the global tilt of the stellar halo, as noted by~\cite{2022arXiv220207662H}.}
\label{table:2}
\end{table}

Lastly, we explore the spatial distributions of the GSE stars in the Test-Set compared to the RV Catalog. The heatmaps in the top row of Fig.~\ref{fig:spatial6D} correspond to the Galactocentric spatial distribution of RV~Catalog stars with $e^\text{truth} > 0.8$.  The bottom row shows the corresponding heatmaps for the Test-Set~Catalog. To quantitatively compare the distribution of GSE candidates, we compute the 3D covariance ellipse of the Galactocentric positions of high-eccentricity stars in each catalog, which has principal axes given by its eigenvectors.  In addition, we project the GSE candidates into 2D Galactocentric planes and obtain the 2D covariance ellipse in each plane; these ellipses are shown for the RV and Test-Set~Catalog in Fig.~\ref{fig:spatial6D}, for stars within 5~kpc of the Sun.   

\begin{figure*}[t!]
\centering
\includegraphics[width=\textwidth]{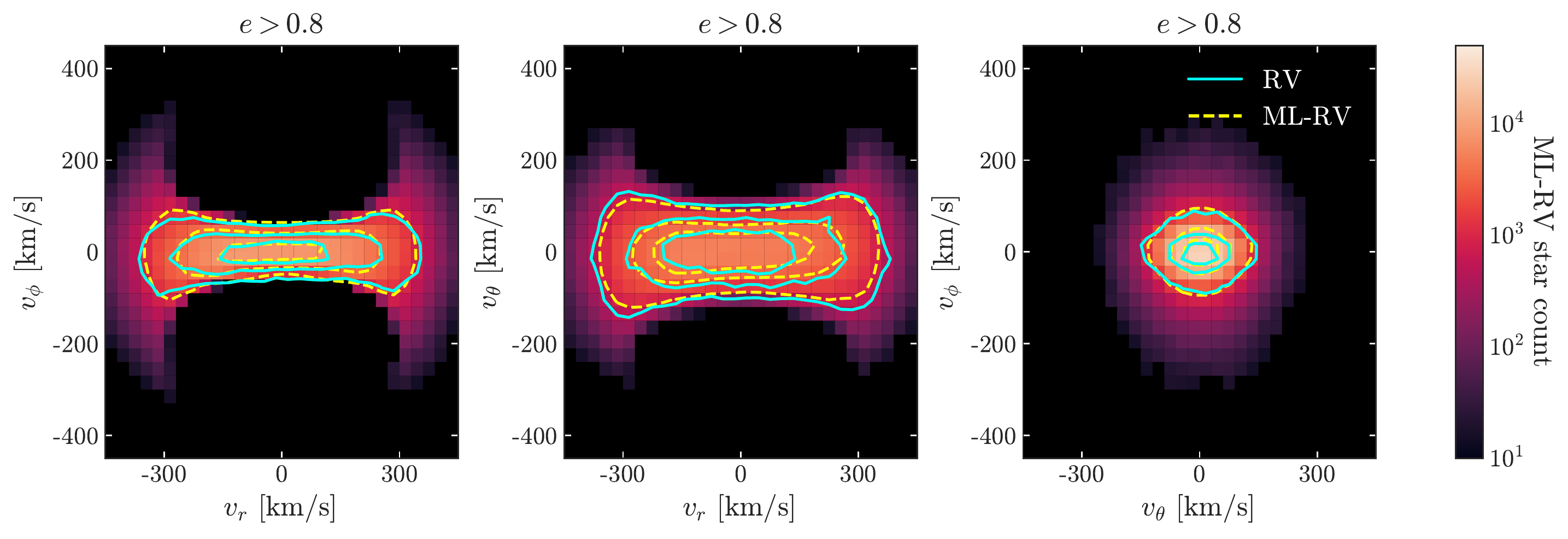}
\caption{Two-dimensional distributions of the Galactocentric velocity components of high-eccentricity stars. The background heatmap shows the network-predicted kinematic distributions of GSE candidate stars ($e^\text{pred} > 0.8$) in the ML-RV Catalog. In this and all following figures, the ML-RV star counts are properly error-sampled, as described in the text. The yellow-dashed contours indicate the location of the 30\%, 60\%, and 90\% containment intervals. For comparison, the solid blue contours indicate the same containment intervals for the GSE candidates ($e^\text{true} > 0.8$) in the RV~Catalog. The GSE's classic ``sausage''-like structure is apparent in the $v_r - v_\phi$ plane for both subsets.}
\label{fig:GSEhighe}
\end{figure*}

Table~\ref{table:2} summarizes two statistics that characterize the spatial properties in each of the catalogs.  The axis ratio is the ratio of the lengths of the principal axes of the 3D covariance ellipse, in decreasing order of length, while the tilt angle is  the angle between the major axis of the 2D covariance ellipse in the $y_\mathrm{gal} - z_\mathrm{gal}$ plane, with respect to the $y_\mathrm{gal}$-axis. For the Test-Set Catalog, the distribution is almost isotropic in the $y_\mathrm{gal} - z_\mathrm{gal}$ plane across the five MC samples. The spatial distribution of Test-Set GSE candidates is  similar to the true distribution, with both distributions being relatively isotropic about the Sun.  We emphasize that the covariance ellipses shown here are used to compare the spatial distribution of GSE candidates between catalogs.  They only apply to local stars within 5~kpc of the Sun and thus do not serve as a characterization of the global stellar halo.

\section{Introducing the ML-RV Catalog}\label{sec:mlcat}

Having validated the neural network's performance in predicting the velocity and spatial distribution of stars with both low- and high-eccentricity orbits, we now apply it to \gaia EDR3 stars with only 5D information. We restrict ourselves to $G \in [12, 17]$ stars to obtain a spatially complete dataset. The predicted line-of-sight velocity and associated uncertainty form what we call the \emph{Machine Learned-Radial Velocity (ML-RV) Catalog}. This catalog is spatially complete, encompassing a total of 91,840,346 stars, more than 20 times larger than the number of equivalent stars in the RV~Catalog.  Table~\ref{table:1} shows how the number of stars in the ML-RV Catalog varies as a function of selection cut.  The majority of stars in the ML-RV Catalog lie within 10~kpc of the Galactic Center, but some are slightly further out, within 20~kpc. 

Both the spatially-complete Test-Set Catalog and the ML-RV Catalog are publicly available for download.  For each catalog, we provide the Gaia EDR3 source ID, predicted line-of-sight velocity $v_\mathrm{los}$ in km~s$^{-1}$, as well as the predicted network uncertainty $\sigma_\mathrm{los}$ in km~s$^{-1}$. \zenodo

\section{Features of GSE in the ML-RV Catalog} \label{sec:enc}

We next identify GSE candidates in the ML-RV~Catalog by looking for stars on high-eccentricity orbits, $e^\mathrm{pred} > 0.8$. This is the first such study of GSE stars in the vast subset of \gaia data with no observed line-of-sight velocities.  Our analysis yields a nearly twentyfold increase in the number of GSE candidate stars as compared to the RV~Catalog.  This section explores the dynamics, abundances, and spatial distribution of these newly identified candidates.  

\begin{figure*}[t!]
\centering
\includegraphics[width=0.9\textwidth]{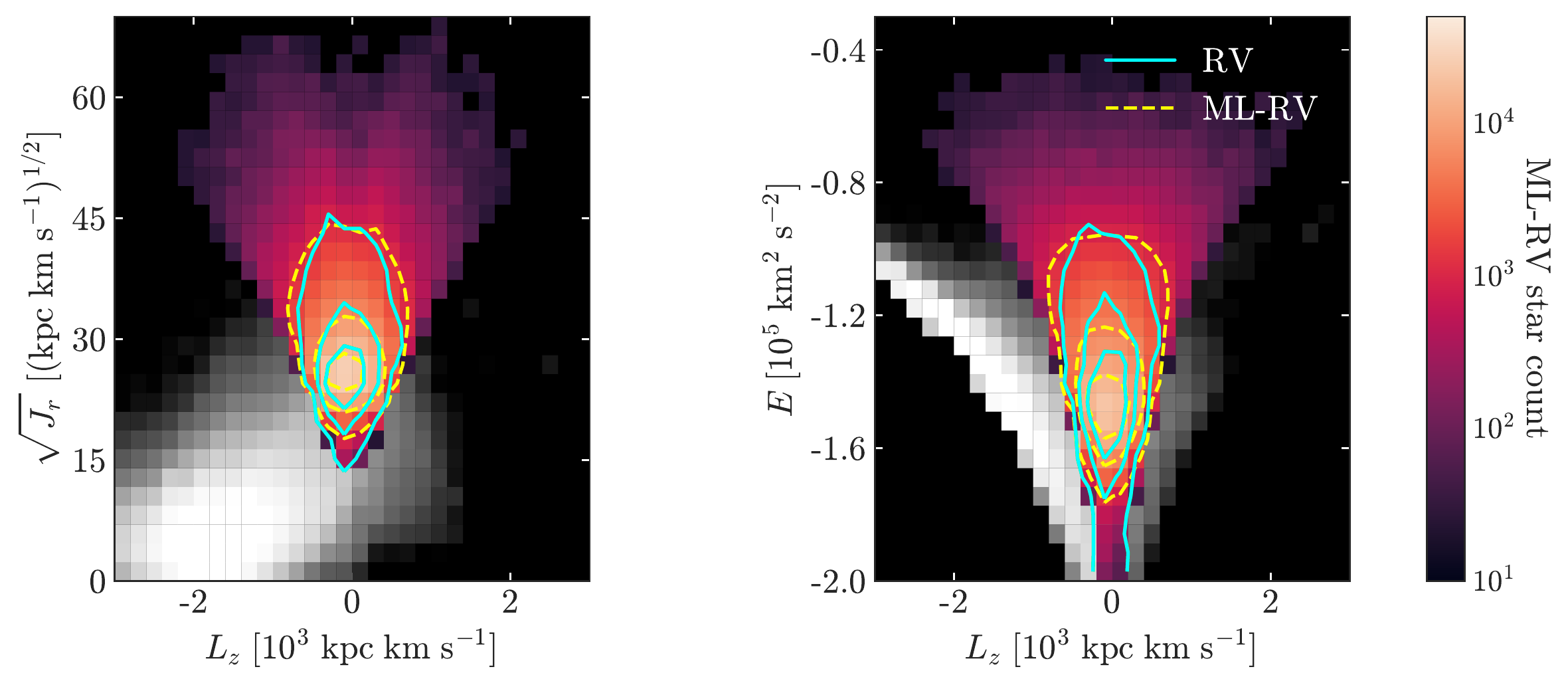}
\caption{Distribution of stars in $E$, $L_z$ and $\sqrt{J_r}$. The pink-hued heatmaps correspond to the integrals of motion for the GSE candidates in the ML-RV~Catalog.  The yellow-dashed contours indicate the location of the 30\%, 60\%, and 90\% containment intervals of these distributions. For comparison, the solid blue contours show the same containment intervals for GSE candidates in the RV~Catalog. The gray-hued (log) density background heatmap shows the  distribution for all stars in the RV~Catalog.}
\label{fig:actions5Dhighe}
\end{figure*}

\subsection{Dynamics}\label{sec:methods_met}

 The Galactocentric spherical velocities of the ML-RV GSE stars are shown in Fig.~\ref{fig:GSEhighe}. The background heatmap corresponds to the error-sampled high-eccentricity ($e^\mathrm{pred} > 0.8$) subset of the  ML-RV Catalog. The dashed yellow lines are the error-sampled 30\%, 60\%, and 90\% containment intervals for the ML-RV Catalog; the solid blue lines are the corresponding truth-level contours for the RV~Catalog.  The inferred velocity distributions of the significantly more numerous ML-RV GSE candidates closely agree with those of the RV~Catalog's GSE stars.

Next, we compare the integrals of motion of the stellar orbits in each catalog, focusing on total orbital energy ($E$), angular momentum in the $z$-direction ($L_z$), and radial action ($\sqrt{J_r}$) for an axisymmetric potential. These quantities are evaluated by integrating orbits in the Milky Way potential given in~\cite{Bovy_2015} using the \texttt{gala} package. Stars from a recent Milky Way merger such as the GSE are expected to show correlation in the integrals of motion  space (see~\cite{helmi_streams_2020} for example).

\begin{figure*}
   \centering
\begin{tabular}{C{\textwidth}}
\includegraphics[width=\textwidth]{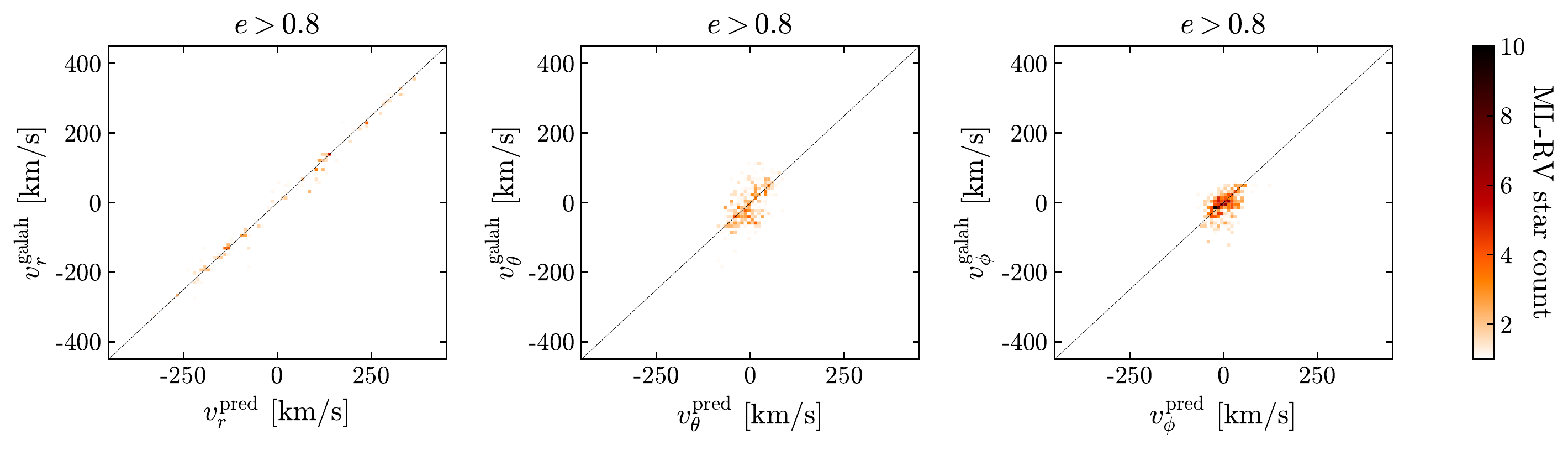}\\
\includegraphics[width=\textwidth]{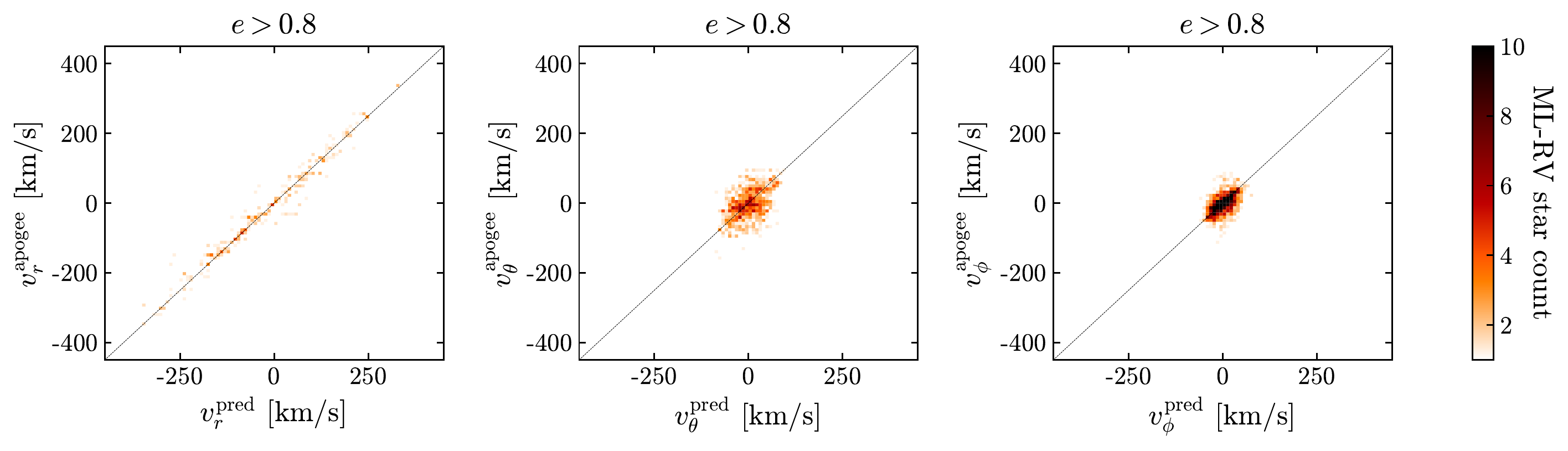}\\
\end{tabular}
    \caption{Comparison of the inferred Galactocentric spherical velocities, $v^{\rm pred}$, in the ML-RV~Catalog with their respective values obtained using line-of-sight velocities from a cross-match to GALAH\texttt{+}~DR3~\emph{(top)} and APOGEE~DR17~\emph{(bottom)}.  The velocities of the low eccentricity stars in the cross-matched ML-RV~Catalog~(not shown here) demonstrate an equally good correspondence with the truth values.}
\label{fig:velcross}
\end{figure*}

Figure~\ref{fig:actions5Dhighe} shows the resulting distributions in the integrals of motion for the GSE stars in the error-sampled ML-RV Catalog, shown as the pink-hued heatmap. The gray-scale background heatmap shows the truth values of the RV~Catalog, which is dominated by the disk, for comparison.  The dashed yellow lines indicate the error-sampled contours for the 30\%, 60\%, and 90\% containment intervals of the ML-RV Catalog stars with $e^{\rm pred} > 0.8$; the solid blue lines are the corresponding truth-level contours for the RV~Catalog stars with $e^{\rm truth}  > 0.8$.

We can compare these distributions with those from the existing literature. \cite{feuillet_selecting_2021} found that stars with $30 \leq \sqrt{J_r}/[(\text{kpc km s}^{-1})^{1/2}] \leq 50$ and $-0.5 \leq L_{z}/[10^3~\text{kpc km s}^{-1}] \leq 0.5$ presented the least-contaminated sample of kinematically-selected GSE stars based on the metallicity distribution. \cite{koppelman_multiple_2019} clustered stars in $E$, $L_z$, $e$, and [Fe/H] to identify GSE stellar candidates. They identified the region occupied predominantly by GSE as ${−1.5 < E/[10^5 \text{ km}^{2} \text{ s}^{−2}] < −1.1}$.  Most of the GSE candidates in the ML-RV Catalog fall reasonably within these energy and angular momentum regions, as seen in Fig.~\ref{fig:actions5Dhighe}. At the 90\% containment interval, we find that the ML-RV GSE sample sits within the ranges: $ 20 \leq \sqrt{J_r}/[(\text{kpc km s}^{-1})^{1/2}]  \leq 44$, $-0.5 \leq L_{z}/[10^3~\text{kpc km s}^{-1}] \leq 0.5$, and ${−1.67 < E/[10^5 \text{ km}^{2} \text{ s}^{−2}] < −1.0}$.  It is important to note that the consistency between our results and those of~\cite{koppelman_multiple_2019, feuillet_selecting_2021} was not guaranteed \emph{a priori}, given that we do not use metallicity information in selecting GSE candidates as they do.

\subsection{Metallicity Distribution}

Using machine-learned line-of-sight velocities to identify GSE candidates in the subset of \gaia data with only 5D astrometry enables us to work with a much larger sample than  otherwise possible by solely relying on cross-matches to spectroscopic surveys.  This benefit does come at the cost of a potentially large network uncertainty on the inferred velocity, which must be properly accounted for when studying the resulting stellar distributions by error sampling.  While the machine learning approach is no match for the gold-standard of spectroscopic observations, it can serve as a complementary tool when complete data is not available.  In this subsection, we focus on the GSE candidates in the ML-RV~Catalog that have cross-matches in GALAH\texttt{+}~DR3~\citep{2020arXiv201102505B} and APOGEE~DR17~\citep{2017AJ....154...94M,queiroz_bulge_2020} to assess the robustness of the network-predicted line-of-sight velocities  and to study their chemical abundance distributions.

There are a total of 689,431~(551,559) stars in the cross-match of GALAH\texttt{+}~(APOGEE) and \gaia EDR3.  The fractions that pass the high-eccentricity ($e>0.8$) selection cut for GSE stars in the RV, Test-Set, and ML-RV~Catalogs are summarized in Tab.~\ref{table:1}.  Figure~\ref{fig:velcross} compares the inferred velocities for the ML-RV high-eccentricity subset with the truth values obtained by combining \emph{Gaia} with the line-of-sight velocities from GALAH\texttt{+}~(top) and APOGEE~(bottow). This serves as an additional cross-check of our machine-learned line-of-sight velocity, using actual measurements that are completely independent of \textit{Gaia}.  There is excellent agreement between the network-inferred and truth values, providing a strong test of the accuracy of the ML-RV velocities.  We have verified that the correspondence is equally good for the low-eccentricity subset as well.

\begin{figure*}[t!]
\centering
\includegraphics[width=0.9\textwidth]{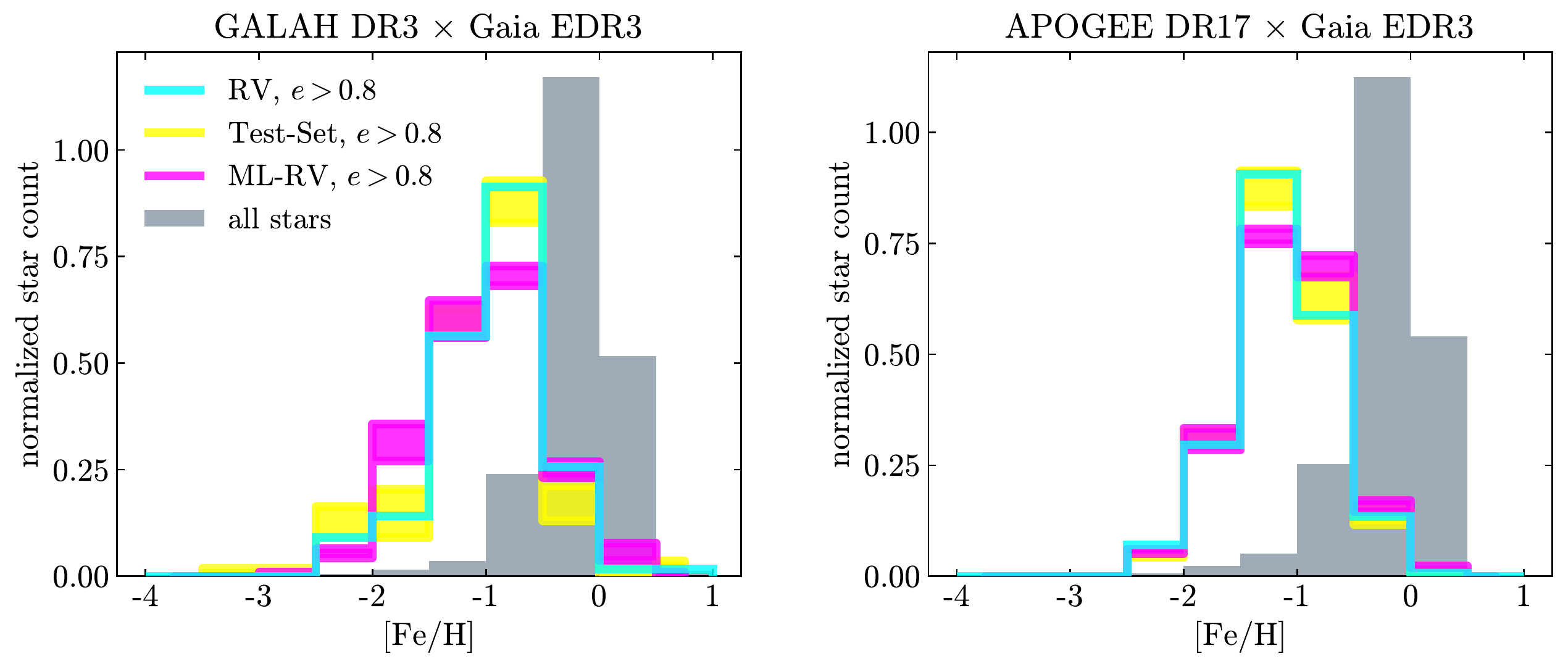}
\caption{The metallicity distributions for GSE candidates in cross-matches of \gaia EDR3 with GALAH\texttt{+}~DR3~\emph{(left)} and APOGEE~DR17~\emph{(right)}.  The distributions for the RV, Test-Set, and ML-RV~Catalogs are shown by the cyan, yellow, and magenta lines, respectively.  The spread in the Test-Set and ML-RV histograms is due to error sampling. The shaded gray histogram corresponds to the distribution of all stars in the cross-matched sample.}
\label{fig:crossmatchhighe}
\end{figure*}

\begin{figure*}[t!]
\centering
\includegraphics[width=0.9\textwidth]{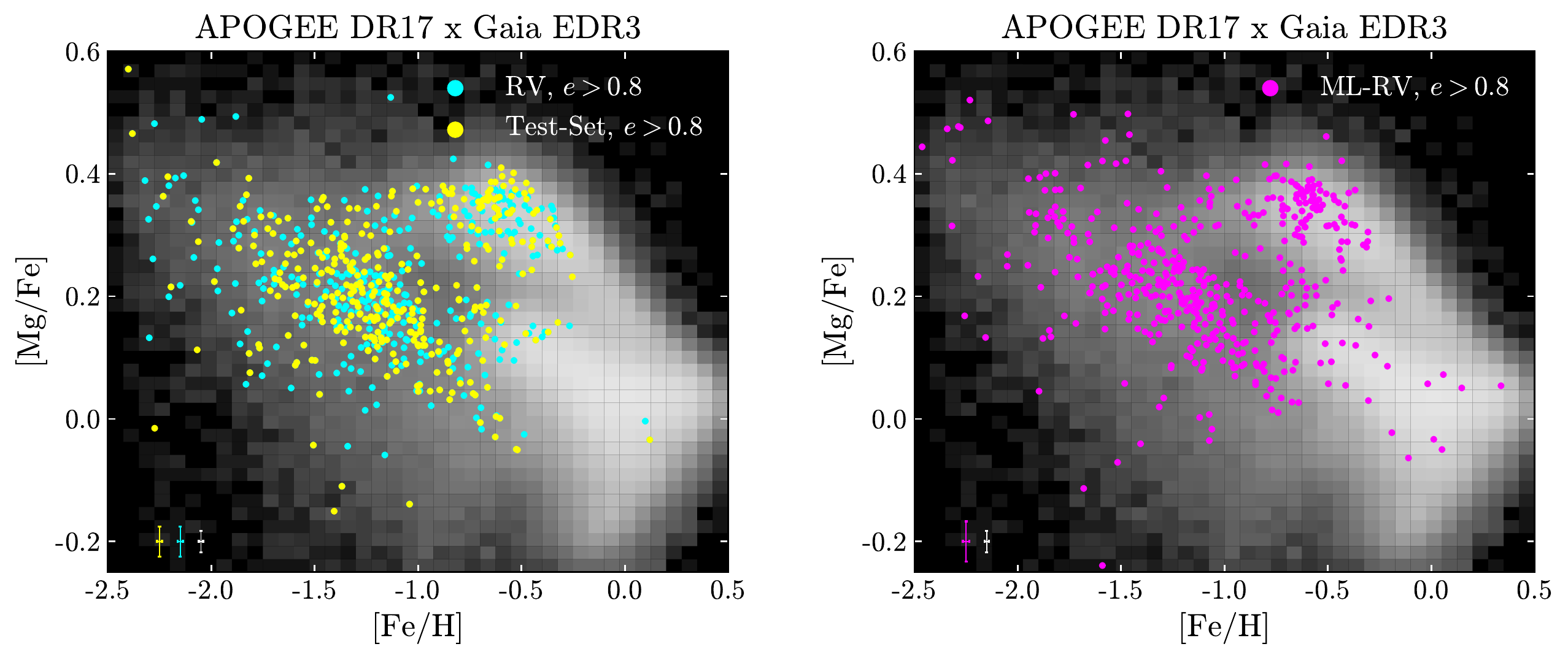}
\caption{Scatter plot of stars in the \gaia EDR3-APOGEE cross-match as a function of $\alpha$-element abundance $\text{[Mg/Fe]}$ and metallicity $\text{[Fe/H]}$.  The results are shown for the GSE candidates in the RV~Catalog~(cyan), as well as a single Monte Carlo iteration of the Test-Set~(yellow) and ML-RV~(magenta) Catalogs. The background grayscale heatmap corresponds to all of the stars in the \gaia EDR3-APOGEE DR17 cross-match~\citep{queiroz_bulge_2020}. The crosses in the bottom left-hand corner of each panel indicate the mean uncertainty in $\text{[Mg/Fe]}$ and $\text{[Fe/H]}$ for each respective sample.}
\label{fig:alpha_fe_match}
\end{figure*}

The left panel of Fig.~\ref{fig:crossmatchhighe} presents the metallicity distribution functions~(MDFs) for the cross-match to GALAH\texttt{+}.  The distributions for high-eccentricity ($e>0.8$) stars in the RV, Test-Set, and ML-RV Catalogs are indicated by the cyan, yellow, and magenta histograms, respectively. The spread in the Test-Set and ML-RV Catalogs is due to error sampling. For comparison, the shaded gray histogram shows the distribution for all stars in the cross-matched sample.  To quantitatively compare the results, we fit the MDFs to a Gaussian distribution. The best-fit mean and standard deviations for the RV and Test-Set distributions are highly consistent with each other with $(\mu, \sigma)= (-0.94, 0.50)$ and $(-1.03, 0.55)$, respectively.\footnote{The quoted mean and standard deviation for the Test-Set and ML-RV Catalogs are averaged across the fits for the five MCs.} The corresponding MDF for the ML-RV GSE candidates is similar: $(\mu, \sigma) = (-1.00, 0.56)$. The right panel of Fig.~\ref{fig:crossmatchhighe} shows the equivalent results for the APOGEE cross-match of high-eccentricity stars.  In this case, the best-fit mean and standard deviations for the RV and Test-Set distributions are $(\mu, \sigma)= (-1.14, 0.45)$ and $(-1.1, 0.43)$, respectively.  The MDF for the ML-RV GSE candidates has $(\mu, \sigma)= (-1.06, 0.47)$.  
These results are consistent with each other, as well as with the GALAH\texttt{+} values presented above.
In Fig.~\ref{fig:alpha_fe_match}, we study the $\alpha$-element abundance [Mg/Fe] of GSE candidate stars from the RV and Test-Set Catalogs~(left panel) and the ML-RV Catalog~(right panel) cross-matched with APOGEE.  The location of a star on the $\text{[Mg/Fe]}-\text{[Fe/H]}$ plane is indicative of its origin in the stellar halo or thin/thick disk.  For example, the grayscale background heatmap in Fig.~\ref{fig:alpha_fe_match} shows all of the stars in the \gaia EDR3-APOGEE~DR17 cross-match~\citep{queiroz_bulge_2020}.  Disk stars correspond to the concentration of stars at metallicities $\text{[Fe/H]} \gtrsim -1.0$, which separate into a high-$\alpha$ and low-$\alpha$ component. Older, accreted stars from mergers typically populate the plane at lower metallicities and extend to higher $\alpha$-abundances---see \emph{e.g.},~\cite{1989ARA&A..27..279W, 1997ARA&A..35..503M, 2009ARA&A..47..371T, 2013A&ARv..21...61R, 2016ARA&A..54..529B} for relevant reviews.

The $\text{[Mg/Fe]}$ distributions of the RV, Test-Set, and ML-RV GSE candidates are consistent with each other.  In all three cases, a significant fraction of the stars are concentrated in a regime that extends from $\text{[Fe/H]} \in [-2.5, -0.75]$ and  $\text{[Mg/Fe]} \in [0.1, 0.4]$, corresponding to an accreted population.  There is also a significant concentration of stars centered at $\text{[Fe/H]} \sim -0.5$ and $\text{[Mg/Fe]} \sim 0.3$, which
would correspond to the high-$\alpha$ disk.  These stars, which have $v_{\phi} \sim 0$, are consistent with the ``Splash" population of \emph{in-situ} stars that may have been perturbed during the GSE merger~\citep{2017ApJ...845..101B, Belokurov_2020}.  Approximately 18\%~(23\%)~(24\%) of the GSE candidates in the  RV~(Test-Set)~(ML-RV) $\times$ APOGEE catalog are in this high-$\alpha$ disk component.

The $\text{[Fe/H]}$ and $\text{[Mg/Fe]}$ distributions that we recover for the GSE candidates in the ML-RV~Catalog are consistent with those previously published~\citep{necib_under_2019,mackereth_origin_2019,2019A&A...632A...4D, bonaca_timing_2020,das_ages_2020,10.1093/mnras/staa1888,10.1093/mnras/staa047,carollo_nature_2021,feuillet_selecting_2021,Hasselquist_2021,2022MNRAS.510.2407B,2022arXiv220404233H}.  This clearly demonstrates that the use of network-inferred velocities---which feeds into the determination of eccentricities---does not bias the resulting abundance distributions. We note that there is also a larger fraction of ML-RV GSE candidates that populate the low-$\alpha$ disk regime, which may suggest some disk contamination in the selection. 

\subsection{Spatial Distribution}

The spatial distribution of GSE debris in the Milky Way is essential for inferring the initial conditions of the  original merger. In the literature, the spatial distribution is still a matter of some debate. By exploiting the excellent sky coverage of RR Lyrae stars,~\cite{10.1093/mnras/stx2819, Iorio_2018} found evidence for a tilted, triaxial stellar halo---likely dominated by GSE debris---within $\sim 30$~kpc of the Galactic Center.  Because the halo is roughly aligned with the Hercules-Aquila Cloud~\citep{Belokurov_2007, 2014MNRAS.440..161S} and the Virgo Overdensity~\citep{Vivas_2001, Newberg_2002, 2006ApJ...636L..97D, Juric_2008, Bonaca_2012}, these authors argue that the overdensities may be associated with the GSE merger---see also~\cite{2019MNRAS.482..921S, 2019ApJ...886...76D}.  \cite{naidu_reconstructing_2021} was able to reproduce these observational results with an isolated $N$-body simulation of a major merger that occurred at infall redshift of $z \sim 2$.  Other studies, however, argue that the tilt in the global halo may be a result of selection biases in RR~Lyrae stars~\citep{helmi_streams_2020}.  As a counterpoint,~\cite{2021A&A...654A..15B} integrate orbits of GSE candidate stars within 2.5~kpc of the Sun and find that the resulting Galactocentric distribution is isotropic.   \cite{2020ApJ...902..119D} show that the Virgo Overdensity and Hercules-Aquila Cloud would only have survived until today if their progenitor fell in $\sim 2.7$~Gyr ago, much more recently than the expectation for the GSE.  However, the orbital integration in both these studies was performed in a spherically symmetric dark matter halo, which has been shown to artificially isotropize stellar distributions~\citep{2022arXiv220207662H}.

The GSE candidate stars studied in this work are concentrated near the Solar position and their spatial distribution is thus representative of the local, not global, stellar halo.  \cite{2022arXiv220207662H} emphasized that the local halo is not a representative sample of the global halo given that  its center is shifted relative to the Galactic Center.  However, as these authors noted, the distribution of the local halo in the $y_{\rm gal} - z_{\rm gal}$ plane may preserve features of the global halo as it is not off-center in this projection.  If so, then the tilt angle of our GSE candidates can potentially be used as an indicator of the global tilt of the stellar halo.  

\begin{figure*}
   \centering
\includegraphics[width=\textwidth]{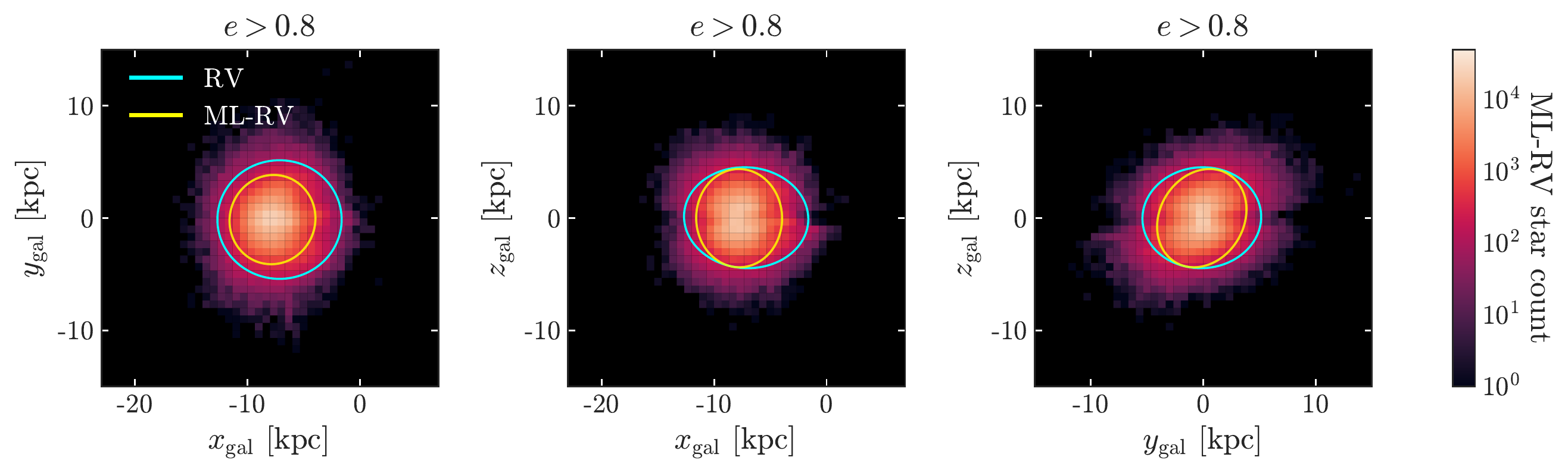}
\caption{Heatmaps corresponding to the Galactocentric spatial distributions for the high-eccentricity stars in the ML-RV~Catalog.  The 2D covariance ellipses for all stars within 5~kpc of the Sun are indicated by the blue~(yellow) lines for the RV~(ML-RV)~Catalog.  There are five essentially indistinguishable
lines shown for the fits to the ML-RV Catalog; each corresponds to a different sample over the network uncertainty. See Tab.~\ref{table:2} for the best-fit axis ratios and tilt angles.}
\label{fig:positions5d}
\end{figure*}

To this end, we now explore the spatial distribution of the GSE candidates in the ML-RV~Catalog, with relevant plots provided in Fig.~\ref{fig:positions5d}.  The background heatmap shows the error-sampled distribution for all stars in the sample. The yellow lines are the 2D covariance ellipses of the spatial coordinates of the candidates within 5~kpc of the Sun.  For comparison, the blue lines show the corresponding 2D covariance ellipses for fits to the RV~Catalog~(same as in Fig.~\ref{fig:spatial6D}).  

As before, we summarize the spatial distribution using the principal axes ratio of the 3D covariance ellipse, as well as the tilt angle of the 2D covariance ellipse in the $y_{\rm gal} - z_{\rm gal}$ plane; these numbers are provided in Tab.~\ref{table:2}. In our validation of the Test-Set Catalog, we found excellent  agreement between truth and predicted elliptical fits.  In both of these cases, the best-fit distribution was essentially isotropic in the heliocentric frame, with no significant tilt in the $y_{\rm gal} - z_{\rm gal}$ plane.  If the local tilt is indicative of the global value, then this result would be consistent with that of~\cite{2021A&A...654A..15B}.  However, the spatial distribution of the ML-RV GSE candidates is very different from that of the RV candidates: we find a significant $\sim 59^\circ$ tilt of these stars above the Galactic plane.  

 We have confirmed that the tilt angles for the RV and Test-Set~Catalogs (see Tab.~\ref{table:2}) remain essentially unchanged when restricting to stars that lie within a heliocentric distance of 2.5~kpc, rather than 5~kpc. In the ML-RV Catalog, we have verified that the tilt is still present, at $69^\circ$, when considering stars within 3.5~kpc; however, it increases to $85^\circ$ (\emph{i.e.}, the ellipse's major axis is almost aligned with the y-axis) for stars within 2.5~kpc. The relative axis ratios remain the same when cutting on distance to the Sun. Moreover, we have verified that the cut on the parallax error does not significantly bias the spatial distribution of stars where these ellipsoidal fits are performed.
Further study will be needed to understand the origin of the observed tilt in the ML-RV GSE sample and whether it is representative of the global halo distribution in the $y_{\rm gal} - z_{\rm gal}$ plane. At the very least, simulations such as that of~\cite{naidu_reconstructing_2021} must also be able to reproduce the local spatial properties of the GSE stars as elucidated by the ML-RV~Catalog.

\section{Conclusion} \label{sec:conclude}

The success of the machine learning approach presented in D21 motivated the application of the method to \gaia data, with the first target goal being the study of the GSE, the most recent major merger in the Milky Way's history. The neural network from D21 takes as input the 5D astrometry of a star and outputs a predicted line-of-sight velocity and associated uncertainty.  In this work, we trained the network on $\sim$ 6.4 million stars from Gaia~EDR3 with complete phase-space information (our ``RV~Catalog”) and ascertained excellent agreement between the true kinematic distributions of these stars and their inferred distributions (our ``Test-Set~Catalog”).  
We then applied the trained network to $\sim$ 92 million stars in \gaia EDR3 without full phase-space information (our ``ML-RV~Catalog''). 

A critical feature of the neural network design is its ability to provide an uncertainty, $\sigma_{\rm los}$, which can be thought of as the network's overall confidence in its velocity prediction.  We explicitly demonstrated that the network uncertainties are nearly Gaussian, so there is a clear understanding of how the predicted line-of-sight velocities deviate from their true values.  Importantly, this Gaussian behavior allows one to create error-sampled distributions on observables of interest, which properly include the network’s uncertainty.

The network's predictive success is quite remarkable given that it is provided with minimal inputs---parallax, two proper motions, and two spatial coordinates.  One can potentially generalize the network to take \emph{e.g.}, action-angle variables, which may enable the exploration of stars out to greater distances in the halo.  However, this would come at the expense of assumptions regarding the Galactic potential that could potentially bias the network output.  One could also explore the possibility of training the network on some combination of \gaia data as well as other spectroscopic surveys, although this would necessitate a careful treatment of potentially  different selection functions between the surveys.

As a first science application of the ML-RV~Catalog, we focused on the identification and characterization of tidal debris of the GSE merger. GSE stars are typically fast moving, retrograde, metal-poor stars on highly eccentric orbits. In this work, we identified GSE candidates using an eccentricity cut of $e > 0.8$.  We confirmed a strong correlation between network-predicted and truth eccentricities, thus justifying the use of an eccentricity selection in the ML-RV~Catalog. We found $\sim 450,000$ GSE candidates, about twenty times more than the number of stars found using the identical selection criteria in the RV~Catalog, where stars have full 6D information.  

The network-predicted distributions of GSE candidates in the 5D \emph{Gaia} catalog have Galactocentric velocities, total orbital energy, radial action, $z$-direction angular momentum, and metallicity-distribution functions that are consistent with the current literature.  In a cross-matched subset of the ML-RV~Catalog, we find that a fraction of the GSE candidates are consistent with the high-$\alpha$ disk and are likely part of the ``Splash," or \emph{in-situ} halo, component.

Finally, we examined the spatial distribution of stars identified as GSE. The spatial distribution of GSE candidates within 5~kpc of the Sun are essentially isotropic in both the RV and Test-Set~Catalog.  The good correspondence between these two catalogs demonstrates that the network does not bias the resulting spatial distributions through its effect on the predicted eccentricity of the stars. 
For ML-RV GSE candidates within a similar distance range, the distribution is anistropic and exhibits a tilt of $59^\circ$ out of the Galactic plane.  Further study is needed to better understand the implications of these findings for the global stellar halo.

The results of this work provide the most extensive mapping of GSE candidates near the Solar position to date, which was made possible by filling in the phase-space of the 5D \gaia EDR3 data.  Any numerical simulation of a GSE-like merger, similar to what was performed by~\cite{naidu_evidence_2020}, must be able to reproduce these results for the local halo.  Through careful comparison of simulation to the spatial and chemodynamic distributions of the GSE candidates in \emph{Gaia}, one can use the simulation to further refine assumptions regarding the initial conditions of the GSE merger, such as its mass, orbital inclination and velocity. Such simulations can be used to investigate, for example, the impact of GSE on the stellar disk, how GSE debris impacts the shape of the dark matter halo and its implication for the rotation of the disk (whether or not it is aligned with the disk). This last point is an especially exciting application as it is directly relevant to mapping the local dark matter velocity distribution in the Milky Way, which has important applications for uncovering the fundamental nature of dark matter in direct dark matter detection experiments~\citep{Necib:2018iwb}.

The ML-RV~Catalog developed in this work can be used for many other applications beyond the study of the GSE.  To this end, we have made the catalog publicly available. \zenodo  One potentially interesting application that has yet to be carefully studied is the use of network-inferred line-of-sight velocities to recover stellar streams.  Identifying cold substructures with inferred kinematics may be a challenge, as the predicted uncertainty may be significantly larger than the dispersion, but the possibility of greatly increasing the statistics available for any analysis of these substructures makes such an approach highly attractive. 

This paper  establishes that machine learning can be used to successfully infer 6D phase-space distributions from 5D astrometry alone.  We outline a systematic approach to train and validate the network in a purely data-driven fashion.  Moreover, through careful treatment of network uncertainties, we explicitly show that network-predicted distributions are robust and interpretable.  When \gaia~DR3 is released, it will include radial velocity measurements for $\sim 30$~million more sources. This wealth of new information will provide an ever-larger training set to use towards improving the predictive power of the network. This work demonstrates that, moving forward, machine learning can serve a critical role in supplementing astrometric and spectroscopic surveys in cases where the available kinematic data is not complete.

\begin{acknowledgments}
The authors would like to thank V.~Belokurov, T.~Cohen, J. Han, Y.~Kahn, L.~Necib, D.~Roberts, and S.~Yaida for fruitful conversations.  ML gratefully acknowledges financial support from the Schmidt DataX Fund at Princeton University made possible through a major gift from the Schmidt Futures Foundation.  BO was supported in part by the U.S. Department of Energy~(DOE) under contract DE-SC0013607 and DE-SC0020223.  HL and ML are supported by the DOE under Award Number DE-SC0007968.  Additionally, HL is supported by NSF grant~PHY-1915409 and the Simons Foundation. AD is supported by NSF~GRFP under Grant No.~DGE-2039656. This work is supported by the National Science Foundation under Cooperative Agreement PHY-2019786 (The NSF AI Institute for Artificial Intelligence and Fundamental Interactions, http://iaifi.org/) and was performed in part at the Aspen Center for Physics, which is supported by NSF grant PHY-1607611.  The work presented in this paper was performed on computational
resources managed and supported by Princeton Research
Computing.  It made use of the \texttt{astropy}~\citep{Robitaille:2013mpa},
\texttt{gala}~\citep{m_price-whelan_gala_2017},
\texttt{corner}~\citep{corner}, \texttt{h5py}~\citep{collette_python_hdf5_2014},  
\texttt{IPython}~\citep{PER-GRA:2007}, 
\texttt{Jupyter}~\citep{Kluyver2016JupyterN}, \texttt{matplotlib}~\citep{Hunter:2007}, 
\texttt{NumPy}~\citep{numpy:2011}, \texttt{pandas}~\citep{mckinney-proc-scipy-2010}, and 
\texttt{SciPy}~\citep{Jones:2001ab} software packages.

This work has made use of data from the European Space Agency (ESA) mission
{\it Gaia} (\url{https://www.cosmos.esa.int/gaia}), processed by the {\it Gaia}
Data Processing and Analysis Consortium (DPAC,
\url{https://www.cosmos.esa.int/web/gaia/dpac/consortium}). Funding for the DPAC
has been provided by national institutions, in particular the institutions
participating in the {\it Gaia} Multilateral Agreement.
\end{acknowledgments}
\clearpage

\bibliography{dataxgaia_part2}{}

\begin{thebibliography}{}
\expandafter\ifx\csname natexlab\endcsname\relax\def\natexlab#1{#1}\fi
\providecommand{\url}[1]{\href{#1}{#1}}
\providecommand{\dodoi}[1]{doi:~\href{http://doi.org/#1}{\nolinkurl{#1}}}
\providecommand{\doeprint}[1]{\href{http://ascl.net/#1}{\nolinkurl{http://ascl.net/#1}}}
\providecommand{\doarXiv}[1]{\href{https://arxiv.org/abs/#1}{\nolinkurl{https://arxiv.org/abs/#1}}}

\bibitem[{{Angus} {et~al.}(2022){Angus}, {Price-Whelan}, {Zinn}, {Bedell},
  {Yuxi}, {Lu}, \& {Foreman-Mackey}}]{2022arXiv220508901A}
{Angus}, R., {Price-Whelan}, A.~M., {Zinn}, J.~C., {et~al.} 2022, arXiv
  e-prints, arXiv:2205.08901.
\newblock \doarXiv{2205.08901}

\bibitem[{{Balbinot} \& {Helmi}(2021)}]{2021A&A...654A..15B}
{Balbinot}, E., \& {Helmi}, A. 2021, \aap, 654, A15,
  \dodoi{10.1051/0004-6361/202141015}

\bibitem[{Batson {et~al.}(2021)Batson, Haaf, Kahn, \&
  Roberts}]{batson_topological_2021}
Batson, J., Haaf, C.~G., Kahn, Y., \& Roberts, D.~A. 2021, Journal of High
  Energy Physics, 2021, 280, \dodoi{10.1007/JHEP04(2021)280}

\bibitem[{Belokurov {et~al.}(2018)Belokurov, Erkal, Evans, Koposov, \&
  Deason}]{belokurov_co-formation_2018}
Belokurov, V., Erkal, D., Evans, N.~W., Koposov, S.~E., \& Deason, A.~J. 2018,
  Monthly Notices of the Royal Astronomical Society, 478, 611,
  \dodoi{10.1093/mnras/sty982}

\bibitem[{Belokurov {et~al.}(2020)Belokurov, Sanders, Fattahi, Smith, Deason,
  Evans, \& Grand}]{Belokurov_2020}
Belokurov, V., Sanders, J.~L., Fattahi, A., {et~al.} 2020, Monthly Notices of
  the Royal Astronomical Society, 494, 3880, \dodoi{10.1093/mnras/staa876}

\bibitem[{Belokurov {et~al.}(2007)Belokurov, Evans, Irwin, Lynden-Bell, Yanny,
  Vidrih, Gilmore, Seabroke, Zucker, Wilkinson, Hewett, Bramich, Fellhauer,
  Newberg, Wyse, Beers, Bell, Barentine, Brinkmann, Cole, Pan, \&
  York}]{Belokurov_2007}
Belokurov, V., Evans, N.~W., Irwin, M.~J., {et~al.} 2007, The Astrophysical
  Journal, 658, 337, \dodoi{10.1086/511302}

\bibitem[{Bird {et~al.}(2019)Bird, Xue, Liu, Shen, Flynn, \&
  Yang}]{bird_anisotropy_2019}
Bird, S.~A., Xue, X.-X., Liu, C., {et~al.} 2019, The Astronomical Journal, 157,
  104, \dodoi{10.3847/1538-3881/aafd2e}

\bibitem[{{Bland-Hawthorn} \& {Gerhard}(2016)}]{2016ARA&A..54..529B}
{Bland-Hawthorn}, J., \& {Gerhard}, O. 2016, \araa, 54, 529,
  \dodoi{10.1146/annurev-astro-081915-023441}

\bibitem[{{Bonaca} {et~al.}(2017){Bonaca}, {Conroy}, {Wetzel}, {Hopkins}, \&
  {Kere{\v{s}}}}]{2017ApJ...845..101B}
{Bonaca}, A., {Conroy}, C., {Wetzel}, A., {Hopkins}, P.~F., \& {Kere{\v{s}}},
  D. 2017, \apj, 845, 101, \dodoi{10.3847/1538-4357/aa7d0c}

\bibitem[{Bonaca {et~al.}(2012)Bonaca, Juri{\'{c}}, Ivezi{\'{c}}, Bizyaev,
  Brewington, Malanushenko, Malanushenko, Oravetz, Pan, Shelden, Simmons, \&
  Snedden}]{Bonaca_2012}
Bonaca, A., Juri{\'{c}}, M., Ivezi{\'{c}}, {\v{Z}}., {et~al.} 2012, The
  Astronomical Journal, 143, 105, \dodoi{10.1088/0004-6256/143/5/105}

\bibitem[{Bonaca {et~al.}(2020)Bonaca, Conroy, Cargile, Naidu, Johnson,
  Zaritsky, Ting, Caldwell, Han, \& van Dokkum}]{bonaca_timing_2020}
Bonaca, A., Conroy, C., Cargile, P.~A., {et~al.} 2020, The Astrophysical
  Journal, 897, L18, \dodoi{10.3847/2041-8213/ab9caa}

\bibitem[{Bovy(2015)}]{Bovy_2015}
Bovy, J. 2015, The Astrophysical Journal Supplement Series, 216, 29,
  \dodoi{10.1088/0067-0049/216/2/29}

\bibitem[{{Buder} {et~al.}(2020){Buder}, {Sharma}, {Kos}, {Amarsi},
  {Nordlander}, {Lind}, {Martell}, {Asplund}, {Bland-Hawthorn}, {Casey}, {De
  Silva}, {D'Orazi}, {Freeman}, {Hayden}, {Lewis}, {Lin}, {Schlesinger},
  {Simpson}, {Stello}, {Zucker}, {Zwitter}, {Beeson}, {Buck}, {Casagrande},
  {Clark}, {Cotar}, {Da Costa}, {de Grijs}, {Feuillet}, {Horner}, {Khanna},
  {Kafle}, {Liu}, {Montet}, {Nandakumar}, {Nataf}, {Ness}, {Spina}, {Traven},
  {Tepper-Garcia}, {Ting}, {Vogrincic}, {Wittenmyer}, {Zerjal}, \& {the GALAH
  collaboration}}]{2020arXiv201102505B}
{Buder}, S., {Sharma}, S., {Kos}, J., {et~al.} 2020, arXiv e-prints,
  arXiv:2011.02505.
\newblock \doarXiv{2011.02505}

\bibitem[{{Buder} {et~al.}(2022){Buder}, {Lind}, {Ness}, {Feuillet}, {Horta},
  {Monty}, {Buck}, {Nordlander}, {Bland-Hawthorn}, {Casey}, {de Silva},
  {D'Orazi}, {Freeman}, {Hayden}, {Kos}, {Martell}, {Lewis}, {Lin},
  {Schlesinger}, {Sharma}, {Simpson}, {Stello}, {Zucker}, {Zwitter},
  {Ciuc{\u{a}}}, {Horner}, {Kobayashi}, {Ting}, {Wyse}, \&
  {Wyse}}]{2022MNRAS.510.2407B}
{Buder}, S., {Lind}, K., {Ness}, M.~K., {et~al.} 2022, \mnras, 510, 2407,
  \dodoi{10.1093/mnras/stab3504}

\bibitem[{Bullock \& Johnston(2005)}]{Bullock_2005}
Bullock, J.~S., \& Johnston, K.~V. 2005, The Astrophysical Journal, 635, 931,
  \dodoi{10.1086/497422}

\bibitem[{Carollo \& Chiba(2021)}]{carollo_nature_2021}
Carollo, D., \& Chiba, M. 2021, The Astrophysical Journal, 908, 191,
  \dodoi{10.3847/1538-4357/abd7a4}

\bibitem[{Collette(2013)}]{collette_python_hdf5_2014}
Collette, A. 2013, Python and HDF5 (O'Reilly)

\bibitem[{Das {et~al.}(2020)Das, Hawkins, \& Jofré}]{das_ages_2020}
Das, P., Hawkins, K., \& Jofré, P. 2020, Monthly Notices of the Royal
  Astronomical Society, 493, 5195, \dodoi{10.1093/mnras/stz3537}

\bibitem[{De~Lucia \& Helmi(2008)}]{DeLucia_2008}
De~Lucia, G., \& Helmi, A. 2008, Monthly Notices of the Royal Astronomical
  Society, 391, 14, \dodoi{10.1111/j.1365-2966.2008.13862.x}

\bibitem[{Deason {et~al.}(2018)Deason, Belokurov, Koposov, \&
  Lancaster}]{deason_apocenter_2018}
Deason, A.~J., Belokurov, V., Koposov, S.~E., \& Lancaster, L. 2018, The
  Astrophysical Journal, 862, L1, \dodoi{10.3847/2041-8213/aad0ee}

\bibitem[{{Di Matteo} {et~al.}(2019){Di Matteo}, {Haywood}, {Lehnert}, {Katz},
  {Khoperskov}, {Snaith}, {G{\'o}mez}, \& {Robichon}}]{2019A&A...632A...4D}
{Di Matteo}, P., {Haywood}, M., {Lehnert}, M.~D., {et~al.} 2019, \aap, 632, A4,
  \dodoi{10.1051/0004-6361/201834929}

\bibitem[{{Donlon} {et~al.}(2021){Donlon}, {Newberg}, {Kim}, \&
  {Lepine}}]{2021arXiv211011465D}
{Donlon}, Thomas, I., {Newberg}, H.~J., {Kim}, B., \& {Lepine}, S. 2021, arXiv
  e-prints, arXiv:2110.11465.
\newblock \doarXiv{2110.11465}

\bibitem[{{Donlon} {et~al.}(2020){Donlon}, {Newberg}, {Sanderson}, \&
  {Widrow}}]{2020ApJ...902..119D}
{Donlon}, Thomas, I., {Newberg}, H.~J., {Sanderson}, R., \& {Widrow}, L.~M.
  2020, \apj, 902, 119, \dodoi{10.3847/1538-4357/abb5f6}

\bibitem[{{Donlon} {et~al.}(2019){Donlon}, {Newberg}, {Weiss}, {Amy}, \&
  {Thompson}}]{2019ApJ...886...76D}
{Donlon}, Thomas, I., {Newberg}, H.~J., {Weiss}, J., {Amy}, P., \& {Thompson},
  J. 2019, \apj, 886, 76, \dodoi{10.3847/1538-4357/ab4f72}

\bibitem[{Dropulic {et~al.}(2021)Dropulic, Ostdiek, Chang, Liu, Cohen, \&
  Lisanti}]{dropulic_machine_2021}
Dropulic, A., Ostdiek, B., Chang, L.~J., {et~al.} 2021, The Astrophysical
  Journal Letters, 915, L14, \dodoi{10.3847/2041-8213/ac09ef}

\bibitem[{{Duffau} {et~al.}(2006){Duffau}, {Zinn}, {Vivas}, {Carraro},
  {M{\'e}ndez}, {Winnick}, \& {Gallart}}]{2006ApJ...636L..97D}
{Duffau}, S., {Zinn}, R., {Vivas}, A.~K., {et~al.} 2006, \apjl, 636, L97,
  \dodoi{10.1086/500130}

\bibitem[{{Fabricius} {et~al.}(2021){Fabricius}, {Luri}, {Arenou}, {Babusiaux},
  {Helmi}, {Muraveva}, {Reyl{\'e}}, {Spoto}, {Vallenari}, {Antoja}, {Balbinot},
  {Barache}, {Bauchet}, {Bragaglia}, {Busonero}, {Cantat-Gaudin}, {Carrasco},
  {Diakit{\'e}}, {Fabrizio}, {Figueras}, {Garcia-Gutierrez}, {Garofalo},
  {Jordi}, {Kervella}, {Khanna}, {Leclerc}, {Licata}, {Lambert}, {Marrese},
  {Masip}, {Ramos}, {Robichon}, {Robin}, {Romero-G{\'o}mez}, {Rubele}, \&
  {Weiler}}]{2021A&A...649A...5F}
{Fabricius}, C., {Luri}, X., {Arenou}, F., {et~al.} 2021, Astronomy and
  Astrophysics, 649, A5, \dodoi{10.1051/0004-6361/202039834}

\bibitem[{Feuillet {et~al.}(2020)Feuillet, Feltzing, Sahlholdt, \&
  Casagrande}]{10.1093/mnras/staa1888}
Feuillet, D.~K., Feltzing, S., Sahlholdt, C.~L., \& Casagrande, L. 2020,
  Monthly Notices of the Royal Astronomical Society, 497, 109,
  \dodoi{10.1093/mnras/staa1888}

\bibitem[{Feuillet {et~al.}(2021)Feuillet, Sahlholdt, Feltzing, \&
  Casagrande}]{feuillet_selecting_2021}
Feuillet, D.~K., Sahlholdt, C.~L., Feltzing, S., \& Casagrande, L. 2021,
  Monthly Notices of the Royal Astronomical Society, 508, 1489,
  \dodoi{10.1093/mnras/stab2614}

\bibitem[{Font {et~al.}(2006)Font, Johnston, Bullock, \& Robertson}]{Font_2006}
Font, A.~S., Johnston, K.~V., Bullock, J.~S., \& Robertson, B.~E. 2006, The
  Astrophysical Journal, 638, 585, \dodoi{10.1086/498970}

\bibitem[{{Forbes}(2020)}]{2020MNRAS.493..847F}
{Forbes}, D.~A. 2020, \mnras, 493, 847, \dodoi{10.1093/mnras/staa245}

\bibitem[{Foreman-Mackey(2016)}]{corner}
Foreman-Mackey, D. 2016, The Journal of Open Source Software, 1, 24,
  \dodoi{10.21105/joss.00024}

\bibitem[{{Gaia Collaboration} {et~al.}(2016){Gaia Collaboration}, {Prusti},
  {de Bruijne}, {Brown}, {Vallenari}, {Babusiaux}, {Bailer-Jones}, {Bastian},
  {Biermann}, {Evans}, {Eyer}, {Jansen}, {Jordi}, {Klioner}, {Lammers},
  {Lindegren}, {Luri}, {Mignard}, {Milligan}, {Panem}, {Poinsignon},
  {Pourbaix}, {Randich}, {Sarri}, {Sartoretti}, {Siddiqui}, {Soubiran},
  {Valette}, {van Leeuwen}, {Walton}, {Aerts}, {Arenou}, {Cropper}, {Drimmel},
  {H{\o}g}, {Katz}, {Lattanzi}, {O'Mullane}, {Grebel}, {Holland}, {Huc},
  {Passot}, {Bramante}, {Cacciari}, {Casta{\~n}eda}, {Chaoul}, {Cheek}, {De
  Angeli}, {Fabricius}, {Guerra}, {Hern{\'a}ndez}, {Jean-Antoine-Piccolo},
  {Masana}, {Messineo}, {Mowlavi}, {Nienartowicz}, {Ord{\'o}{\~n}ez-Blanco},
  {Panuzzo}, {Portell}, {Richards}, {Riello}, {Seabroke}, {Tanga},
  {Th{\'e}venin}, {Torra}, {Els}, {Gracia-Abril}, {Comoretto},
  {Garcia-Reinaldos}, {Lock}, {Mercier}, {Altmann}, {Andrae}, {Astraatmadja},
  {Bellas-Velidis}, {Benson}, {Berthier}, {Blomme}, {Busso}, {Carry},
  {Cellino}, {Clementini}, {Cowell}, {Creevey}, {Cuypers}, {Davidson}, {De
  Ridder}, {de Torres}, {Delchambre}, {Dell'Oro}, {Ducourant}, {Fr{\'e}mat},
  {Garc{\'\i}a-Torres}, {Gosset}, {Halbwachs}, {Hambly}, {Harrison}, {Hauser},
  {Hestroffer}, {Hodgkin}, {Huckle}, {Hutton}, {Jasniewicz}, {Jordan},
  {Kontizas}, {Korn}, {Lanzafame}, {Manteiga}, {Moitinho}, {Muinonen},
  {Osinde}, {Pancino}, {Pauwels}, {Petit}, {Recio-Blanco}, {Robin}, {Sarro},
  {Siopis}, {Smith}, {Smith}, {Sozzetti}, {Thuillot}, {van Reeven}, {Viala},
  {Abbas}, {Abreu Aramburu}, {Accart}, {Aguado}, {Allan}, {Allasia},
  {Altavilla}, {{\'A}lvarez}, {Alves}, {Anderson}, {Andrei}, {Anglada Varela},
  {Antiche}, {Antoja}, {Ant{\'o}n}, {Arcay}, {Atzei}, {Ayache}, {Bach},
  {Baker}, {Balaguer-N{\'u}{\~n}ez}, {Barache}, {Barata}, {Barbier}, {Barblan},
  {Baroni}, {Barrado y Navascu{\'e}s}, {Barros}, {Barstow}, {Becciani},
  {Bellazzini}, {Bellei}, {Bello Garc{\'\i}a}, {Belokurov}, {Bendjoya},
  {Berihuete}, {Bianchi}, {Bienaym{\'e}}, {Billebaud}, {Blagorodnova},
  {Blanco-Cuaresma}, {Boch}, {Bombrun}, {Borrachero}, {Bouquillon}, {Bourda},
  {Bouy}, {Bragaglia}, {Breddels}, {Brouillet}, {Br{\"u}semeister},
  {Bucciarelli}, {Budnik}, {Burgess}, {Burgon}, {Burlacu}, {Busonero}, {Buzzi},
  {Caffau}, {Cambras}, {Campbell}, {Cancelliere}, {Cantat-Gaudin}, {Carlucci},
  {Carrasco}, {Castellani}, {Charlot}, {Charnas}, {Charvet}, {Chassat},
  {Chiavassa}, {Clotet}, {Cocozza}, {Collins}, {Collins}, {Costigan}, {Crifo},
  {Cross}, {Crosta}, {Crowley}, {Dafonte}, {Damerdji}, {Dapergolas}, {David},
  {David}, {De Cat}, {de Felice}, {de Laverny}, {De Luise}, {De March}, {de
  Martino}, {de Souza}, {Debosscher}, {del Pozo}, {Delbo}, {Delgado},
  {Delgado}, {di Marco}, {Di Matteo}, {Diakite}, {Distefano}, {Dolding}, {Dos
  Anjos}, {Drazinos}, {Dur{\'a}n}, {Dzigan}, {Ecale}, {Edvardsson}, {Enke},
  {Erdmann}, {Escolar}, {Espina}, {Evans}, {Eynard Bontemps}, {Fabre},
  {Fabrizio}, {Faigler}, {Falc{\~a}o}, {Farr{\`a}s Casas}, {Faye}, {Federici},
  {Fedorets}, {Fern{\'a}ndez-Hern{\'a}ndez}, {Fernique}, {Fienga}, {Figueras},
  {Filippi}, {Findeisen}, {Fonti}, {Fouesneau}, {Fraile}, {Fraser}, {Fuchs},
  {Furnell}, {Gai}, {Galleti}, {Galluccio}, {Garabato}, {Garc{\'\i}a-Sedano},
  {Gar{\'e}}, {Garofalo}, {Garralda}, {Gavras}, {Gerssen}, {Geyer}, {Gilmore},
  {Girona}, {Giuffrida}, {Gomes}, {Gonz{\'a}lez-Marcos},
  {Gonz{\'a}lez-N{\'u}{\~n}ez}, {Gonz{\'a}lez-Vidal}, {Granvik}, {Guerrier},
  {Guillout}, {Guiraud}, {G{\'u}rpide}, {Guti{\'e}rrez-S{\'a}nchez}, {Guy},
  {Haigron}, {Hatzidimitriou}, {Haywood}, {Heiter}, {Helmi}, {Hobbs},
  {Hofmann}, {Holl}, {Holland}, {Hunt}, {Hypki}, {Icardi}, {Irwin}, {Jevardat
  de Fombelle}, {Jofr{\'e}}, {Jonker}, {Jorissen}, {Julbe}, {Karampelas},
  {Kochoska}, {Kohley}, {Kolenberg}, {Kontizas}, {Koposov}, {Kordopatis},
  {Koubsky}, {Kowalczyk}, {Krone-Martins}, {Kudryashova}, {Kull}, {Bachchan},
  {Lacoste-Seris}, {Lanza}, {Lavigne}, {Le Poncin-Lafitte}, {Lebreton},
  {Lebzelter}, {Leccia}, {Leclerc}, {Lecoeur-Taibi}, {Lemaitre}, {Lenhardt},
  {Leroux}, {Liao}, {Licata}, {Lindstr{\o}m}, {Lister}, {Livanou}, {Lobel},
  {L{\"o}ffler}, {L{\'o}pez}, {Lopez-Lozano}, {Lorenz}, {Loureiro},
  {MacDonald}, {Magalh{\~a}es Fernandes}, {Managau}, {Mann}, {Mantelet},
  {Marchal}, {Marchant}, {Marconi}, {Marie}, {Marinoni}, {Marrese},
  {Marschalk{\'o}}, {Marshall}, {Mart{\'\i}n-Fleitas}, {Martino}, {Mary},
  {Matijevi{\v{c}}}, {Mazeh}, {McMillan}, {Messina}, {Mestre}, {Michalik},
  {Millar}, {Miranda}, {Molina}, {Molinaro}, {Molinaro}, {Moln{\'a}r},
  {Moniez}, {Montegriffo}, {Monteiro}, {Mor}, {Mora}, {Morbidelli}, {Morel},
  {Morgenthaler}, {Morley}, {Morris}, {Mulone}, {Muraveva}, {Musella},
  {Narbonne}, {Nelemans}, {Nicastro}, {Noval}, {Ord{\'e}novic},
  {Ordieres-Mer{\'e}}, {Osborne}, {Pagani}, {Pagano}, {Pailler}, {Palacin},
  {Palaversa}, {Parsons}, {Paulsen}, {Pecoraro}, {Pedrosa}, {Pentik{\"a}inen},
  {Pereira}, {Pichon}, {Piersimoni}, {Pineau}, {Plachy}, {Plum}, {Poujoulet},
  {Pr{\v{s}}a}, {Pulone}, {Ragaini}, {Rago}, {Rambaux}, {Ramos-Lerate},
  {Ranalli}, {Rauw}, {Read}, {Regibo}, {Renk}, {Reyl{\'e}}, {Ribeiro},
  {Rimoldini}, {Ripepi}, {Riva}, {Rixon}, {Roelens}, {Romero-G{\'o}mez},
  {Rowell}, {Royer}, {Rudolph}, {Ruiz-Dern}, {Sadowski}, {Sagrist{\`a}
  Sell{\'e}s}, {Sahlmann}, {Salgado}, {Salguero}, {Sarasso}, {Savietto},
  {Schnorhk}, {Schultheis}, {Sciacca}, {Segol}, {Segovia}, {Segransan},
  {Serpell}, {Shih}, {Smareglia}, {Smart}, {Smith}, {Solano}, {Solitro},
  {Sordo}, {Soria Nieto}, {Souchay}, {Spagna}, {Spoto}, {Stampa}, {Steele},
  {Steidelm{\"u}ller}, {Stephenson}, {Stoev}, {Suess}, {S{\"u}veges}, {Surdej},
  {Szabados}, {Szegedi-Elek}, {Tapiador}, {Taris}, {Tauran}, {Taylor},
  {Teixeira}, {Terrett}, {Tingley}, {Trager}, {Turon}, {Ulla}, {Utrilla},
  {Valentini}, {van Elteren}, {Van Hemelryck}, {van Leeuwen}, {Varadi},
  {Vecchiato}, {Veljanoski}, {Via}, {Vicente}, {Vogt}, {Voss}, {Votruba},
  {Voutsinas}, {Walmsley}, {Weiler}, {Weingrill}, {Werner}, {Wevers},
  {Whitehead}, {Wyrzykowski}, {Yoldas}, {{\v{Z}}erjal}, {Zucker}, {Zurbach},
  {Zwitter}, {Alecu}, {Allen}, {Allende Prieto}, {Amorim},
  {Anglada-Escud{\'e}}, {Arsenijevic}, {Azaz}, {Balm}, {Beck}, {Bernstein},
  {Bigot}, {Bijaoui}, {Blasco}, {Bonfigli}, {Bono}, {Boudreault}, {Bressan},
  {Brown}, {Brunet}, {Bunclark}, {Buonanno}, {Butkevich}, {Carret}, {Carrion},
  {Chemin}, {Ch{\'e}reau}, {Corcione}, {Darmigny}, {de Boer}, {de Teodoro}, {de
  Zeeuw}, {Delle Luche}, {Domingues}, {Dubath}, {Fodor}, {Fr{\'e}zouls},
  {Fries}, {Fustes}, {Fyfe}, {Gallardo}, {Gallegos}, {Gardiol}, {Gebran},
  {Gomboc}, {G{\'o}mez}, {Grux}, {Gueguen}, {Heyrovsky}, {Hoar}, {Iannicola},
  {Isasi Parache}, {Janotto}, {Joliet}, {Jonckheere}, {Keil}, {Kim},
  {Klagyivik}, {Klar}, {Knude}, {Kochukhov}, {Kolka}, {Kos}, {Kutka}, {Lainey},
  {LeBouquin}, {Liu}, {Loreggia}, {Makarov}, {Marseille}, {Martayan},
  {Martinez-Rubi}, {Massart}, {Meynadier}, {Mignot}, {Munari}, {Nguyen},
  {Nordlander}, {Ocvirk}, {O'Flaherty}, {Olias Sanz}, {Ortiz}, {Osorio},
  {Oszkiewicz}, {Ouzounis}, {Palmer}, {Park}, {Pasquato}, {Peltzer}, {Peralta},
  {P{\'e}turaud}, {Pieniluoma}, {Pigozzi}, {Poels}, {Prat}, {Prod'homme},
  {Raison}, {Rebordao}, {Risquez}, {Rocca-Volmerange}, {Rosen}, {Ruiz-Fuertes},
  {Russo}, {Sembay}, {Serraller Vizcaino}, {Short}, {Siebert}, {Silva},
  {Sinachopoulos}, {Slezak}, {Soffel}, {Sosnowska}, {Strai{\v{z}}ys}, {ter
  Linden}, {Terrell}, {Theil}, {Tiede}, {Troisi}, {Tsalmantza}, {Tur},
  {Vaccari}, {Vachier}, {Valles}, {Van Hamme}, {Veltz}, {Virtanen}, {Wallut},
  {Wichmann}, {Wilkinson}, {Ziaeepour}, \& {Zschocke}}]{2016A&A...595A...1G}
{Gaia Collaboration}, {Prusti}, T., {de Bruijne}, J.~H.~J., {et~al.} 2016,
  \aap, 595, A1, \dodoi{10.1051/0004-6361/201629272}

\bibitem[{{Gaia Collaboration} {et~al.}(2018){Gaia Collaboration}, {Katz, D.},
  {Antoja, T.}, {Romero-G{\'o}mez, M.}, {Drimmel, R.}, {Reyl{\'e}, C.},
  {Seabroke, G. M.}, {Soubiran, C.}, {Babusiaux, C.}, {Di Matteo, P.},
  {Figueras, F.}, {Poggio, E.}, {Robin, A. C.}, {Evans, D. W.}, {Brown, A. G.
  A.}, {Vallenari, A.}, {Prusti, T.}, {de Bruijne, J. H. J.}, {Bailer-Jones, C.
  A. L.}, {Biermann, M.}, {Eyer, L.}, {Jansen, F.}, {Jordi, C.}, {Klioner, S.
  A.}, {Lammers, U.}, {Lindegren, L.}, {Luri, X.}, {Mignard, F.}, {Panem, C.},
  {Pourbaix, D.}, {Randich, S.}, {Sartoretti, P.}, {Siddiqui, H. I.}, {van
  Leeuwen, F.}, {Walton, N. A.}, {Arenou, F.}, {Bastian, U.}, {Cropper, M.},
  {Lattanzi, M. G.}, {Bakker, J.}, {Cacciari, C.}, {Casta n, J.}, {Chaoul, L.},
  {Cheek, N.}, {De Angeli, F.}, {Fabricius, C.}, {Guerra, R.}, {Holl, B.},
  {Masana, E.}, {Messineo, R.}, {Mowlavi, N.}, {Nienartowicz, K.}, {Panuzzo,
  P.}, {Portell, J.}, {Riello, M.}, {Tanga, P.}, {Th{\'e}venin, F.},
  {Gracia-Abril, G.}, {Comoretto, G.}, {Garcia-Reinaldos, M.}, {Teyssier, D.},
  {Altmann, M.}, {Andrae, R.}, {Audard, M.}, {Bellas-Velidis, I.}, {Benson,
  K.}, {Berthier, J.}, {Blomme, R.}, {Burgess, P.}, {Busso, G.}, {Carry, B.},
  {Cellino, A.}, {Clementini, G.}, {Clotet, M.}, {Creevey, O.}, {Davidson, M.},
  {De Ridder, J.}, {Delchambre, L.}, {Dell{\textbackslash}'Oro, A.},
  {Ducourant, C.}, {Fern{\'a}ndez-Hern{\'a}ndez, J.}, {Fouesneau, M.},
  {Fr{\'e}mat, Y.}, {Galluccio, L.}, {Garc{\'\i}a-Torres, M.},
  {Gonz{\'a}lez-N{\'u}{\~n}ez, J.}, {Gonz{\'a}lez-Vidal, J. J.}, {Gosset, E.},
  {Guy, L. P.}, {Halbwachs, J.-L.}, {Hambly, N. C.}, {Harrison, D. L.},
  {Hern{\'a}ndez, J.}, {Hestroffer, D.}, {Hodgkin, S. T.}, {Hutton, A.},
  {Jasniewicz, G.}, {Jean-Antoine-Piccolo, A.}, {Jordan, S.}, {Korn, A. J.},
  {Krone-Martins, A.}, {Lanzafame, A. C.}, {Lebzelter, T.}, {L{\"o}ffler, W.},
  {Manteiga, M.}, {Marrese, P. M.}, {Mart{\'\i}n-Fleitas, J. M.}, {Moitinho,
  A.}, {Mora, A.}, {Muinonen, K.}, {Osinde, J.}, {Pancino, E.}, {Pauwels, T.},
  {Petit, J.-M.}, {Recio-Blanco, A.}, {Richards, P. J.}, {Rimoldini, L.},
  {Sarro, L. M.}, {Siopis, C.}, {Smith, M.}, {Sozzetti, A.}, {S{\"u}veges, M.},
  {Torra, J.}, {van Reeven, W.}, {Abbas, U.}, {Abreu Aramburu, A.}, {Accart,
  S.}, {Aerts, C.}, {Altavilla, G.}, {{\'A}lvarez, M. A.}, {Alvarez, R.},
  {Alves, J.}, {Anderson, R. I.}, {Andrei, A. H.}, {Anglada Varela, E.},
  {Antiche, E.}, {Arcay, B.}, {Astraatmadja, T. L.}, {Bach, N.}, {Baker, S.
  G.}, {Balaguer-N{\'u}{\~n}ez, L.}, {Balm, P.}, {Barache, C.}, {Barata, C.},
  {Barbato, D.}, {Barblan, F.}, {Barklem, P. S.}, {Barrado, D.}, {Barros, M.},
  {Barstow, M. A.}, {Bartholom{\'e} Mu{\~n}oz, L.}, {Bassilana, J.-L.},
  {Becciani, U.}, {Bellazzini, M.}, {Berihuete, A.}, {Bertone, S.}, {Bianchi,
  L.}, {Bienaym{\'e}, O.}, {Blanco-Cuaresma, S.}, {Boch, T.}, {Boeche, C.},
  {Bombrun, A.}, {Borrachero, R.}, {Bossini, D.}, {Bouquillon, S.}, {Bourda,
  G.}, {Bragaglia, A.}, {Bramante, L.}, {Breddels, M. A.}, {Bressan, A.},
  {Brouillet, N.}, {Br{\"u}semeister, T.}, {Brugaletta, E.}, {Bucciarelli, B.},
  {Burlacu, A.}, {Busonero, D.}, {Butkevich, A. G.}, {Buzzi, R.}, {Caffau, E.},
  {Cancelliere, R.}, {Cannizzaro, G.}, {Cantat-Gaudin, T.}, {Carballo, R.},
  {Carlucci, T.}, {Carrasco, J. M.}, {Casamiquela, L.}, {Castellani, M.},
  {Castro-Ginard, A.}, {Charlot, P.}, {Chemin, L.}, {Chiavassa, A.}, {Cocozza,
  G.}, {Costigan, G.}, {Cowell, S.}, {Crifo, F.}, {Crosta, M.}, {Crowley, C.},
  {Cuypers+, J.}, {Dafonte, C.}, {Damerdji, Y.}, {Dapergolas, A.}, {David, P.},
  {David, M.}, {de Laverny, P.}, {De Luise, F.}, {De March, R.}, {de Souza,
  R.}, {de Torres, A.}, {Debosscher, J.}, {del Pozo, E.}, {Delbo, M.},
  {Delgado, A.}, {Delgado, H. E.}, {Diakite, S.}, {Diener, C.}, {Distefano,
  E.}, {Dolding, C.}, {Drazinos, P.}, {Dur{\'a}n, J.}, {Edvardsson, B.}, {Enke,
  H.}, {Eriksson, K.}, {Esquej, P.}, {Eynard Bontemps, G.}, {Fabre, C.},
  {Fabrizio, M.}, {Faigler, S.}, {Falc a, A. J.}, {Farr{\`a}s Casas, M.},
  {Federici, L.}, {Fedorets, G.}, {Fernique, P.}, {Filippi, F.}, {Findeisen,
  K.}, {Fonti, A.}, {Fraile, E.}, {Fraser, M.}, {Fr{\'e}zouls, B.}, {Gai, M.},
  {Galleti, S.}, {Garabato, D.}, {Garc{\'\i}a-Sedano, F.}, {Garofalo, A.},
  {Garralda, N.}, {Gavel, A.}, {Gavras, P.}, {Gerssen, J.}, {Geyer, R.},
  {Giacobbe, P.}, {Gilmore, G.}, {Girona, S.}, {Giuffrida, G.}, {Glass, F.},
  {Gomes, M.}, {Granvik, M.}, {Gueguen, A.}, {Guerrier, A.}, {Guiraud, J.},
  {Guti{\'e}, R.}, {Haigron, R.}, {Hatzidimitriou, D.}, {Hauser, M.}, {Haywood,
  M.}, {Heiter, U.}, {Helmi, A.}, {Heu, J.}, {Hilger, T.}, {Hobbs, D.},
  {Hofmann, W.}, {Holland, G.}, {Huckle, H. E.}, {Hypki, A.}, {Icardi, V.},
  {Jan{\ss}en, K.}, {Jevardat de Fombelle, G.}, {Jonker, P. G.}, {Juh{\'a}sz,
  {\'A}. L.}, {Julbe, F.}, {Karampelas, A.}, {Kewley, A.}, {Klar, J.},
  {Kochoska, A.}, {Kohley, R.}, {Kolenberg, K.}, {Kontizas, M.}, {Kontizas,
  E.}, {Koposov, S. E.}, {Kordopatis, G.}, {Kostrzewa-Rutkowska, Z.}, {Koubsky,
  P.}, {Lambert, S.}, {Lanza, A. F.}, {Lasne, Y.}, {Lavigne, J.-B.}, {Le
  Fustec, Y.}, {Le Poncin-Lafitte, C.}, {Lebreton, Y.}, {Leccia, S.}, {Leclerc,
  N.}, {Lecoeur-Taibi, I.}, {Lenhardt, H.}, {Leroux, F.}, {Liao, S.}, {Licata,
  E.}, {Lindstr{\o}m, H. E. P.}, {Lister, T. A.}, {Livanou, E.}, {Lobel, A.},
  {L{\'o}pez, M.}, {Managau, S.}, {Mann, R. G.}, {Mantelet, G.}, {Marchal, O.},
  {Marchant, J. M.}, {Marconi, M.}, {Marinoni, S.}, {Marschalk{\'o}, G.},
  {Marshall, D. J.}, {Martino, M.}, {Marton, G.}, {Mary, N.}, {Massari, D.},
  {Matijevic, G.}, {Mazeh, T.}, {McMillan, P. J.}, {Messina, S.}, {Michalik,
  D.}, {Millar, N. R.}, {Molina, D.}, {Molinaro, R.}, {Moln{\'a}r, L.},
  {Montegriffo, P.}, {Mor, R.}, {Morbidelli, R.}, {Morel, T.}, {Morris, D.},
  {Mulone, A. F.}, {Muraveva, T.}, {Musella, I.}, {Nelemans, G.}, {Nicastro,
  L.}, {Noval, L.}, {O{\textbackslash}'Mullane, W.}, {Ord{\'e}novic, C.},
  {Ord{\'o}{\~n}ez-Blanco, D.}, {Osborne, P.}, {Pagani, C.}, {Pagano, I.},
  {Pailler, F.}, {Palacin, H.}, {Palaversa, L.}, {Panahi, A.}, {Pawlak, M.},
  {Piersimoni, A. M.}, {Pineau, F.-X.}, {Plachy, E.}, {Plum, G.}, {Poujoulet,
  E.}, {Prsa, A.}, {Pulone, L.}, {Racero, E.}, {Ragaini, S.}, {Rambaux, N.},
  {Ramos-Lerate, M.}, {Regibo, S.}, {Riclet, F.}, {Ripepi, V.}, {Riva, A.},
  {Rivard, A.}, {Rixon, G.}, {Roegiers, T.}, {Roelens, M.}, {Rowell, N.},
  {Royer, F.}, {Ruiz-Dern, L.}, {Sadowski, G.}, {Sagrist{\`a} Sell{\'e}s, T.},
  {Sahlmann, J.}, {Salgado, J.}, {Salguero, E.}, {Sanna, N.}, {Santana-Ros,
  T.}, {Sarasso, M.}, {Savietto, H.}, {Schultheis, M.}, {Sciacca, E.}, {Segol,
  M.}, {Segovia, J. C.}, {S{\'e}gransan, D.}, {Shih, I-C.}, {Siltala, L.},
  {Silva, A. F.}, {Smart, R. L.}, {Smith, K. W.}, {Solano, E.}, {Solitro, F.},
  {Sordo, R.}, {Soria Nieto, S.}, {Souchay, J.}, {Spagna, A.}, {Spoto, F.},
  {Stampa, U.}, {Steele, I. A.}, {Steidelm{\"u}ller, H.}, {Stephenson, C. A.},
  {Stoev, H.}, {Suess, F. F.}, {Surdej, J.}, {Szabados, L.}, {Szegedi-Elek,
  E.}, {Tapiador, D.}, {Taris, F.}, {Tauran, G.}, {Taylor, M. B.}, {Teixeira,
  R.}, {Terrett, D.}, {Teyssandier, P.}, {Thuillot, W.}, {Titarenko, A.},
  {Torra Clotet, F.}, {Turon, C.}, {Ulla, A.}, {Utrilla, E.}, {Uzzi, S.},
  {Vaillant, M.}, {Valentini, G.}, {Valette, V.}, {van Elteren, A.}, {Van
  Hemelryck, E.}, {van Leeuwen, M.}, {Vaschetto, M.}, {Vecchiato, A.},
  {Veljanoski, J.}, {Viala, Y.}, {Vicente, D.}, {Vogt, S.}, {von Essen, C.},
  {Voss, H.}, {Votruba, V.}, {Voutsinas, S.}, {Walmsley, G.}, {Weiler, M.},
  {Wertz, O.}, {Wevers, T.}, {Wyrzykowski, L.}, {Yoldas, A.}, {Zerjal, M.},
  {Ziaeepour, H.}, {Zorec, J.}, {Zschocke, S.}, {Zucker, S.}, {Zurbach, C.}, \&
  {Zwitter, T.}}]{gaia_collaboration_gaia_2018}
{Gaia Collaboration}, {Katz, D.}, {Antoja, T.}, {et~al.} 2018, Astronomy and
  Astrophysics, 616, A11, \dodoi{10.1051/0004-6361/201832865}

\bibitem[{{Gaia Collaboration} {et~al.}(2021){Gaia Collaboration}, Brown,
  Vallenari, Prusti, de~Bruijne, Babusiaux, Biermann, Creevey, Evans, Eyer,
  Hutton, Jansen, Jordi, Klioner, Lammers, Lindegren, Luri, Mignard, Panem,
  Pourbaix, Randich, Sartoretti, Soubiran, Walton, Arenou, Bailer-Jones,
  Bastian, Cropper, Drimmel, Katz, Lattanzi, van Leeuwen, Bakker, Cacciari,
  Castañeda, De~Angeli, Ducourant, Fabricius, Fouesneau, Frémat, Guerra,
  Guerrier, Guiraud, Jean-Antoine~Piccolo, Masana, Messineo, Mowlavi, Nicolas,
  Nienartowicz, Pailler, Panuzzo, Riclet, Roux, Seabroke, Sordo, Tanga,
  Thévenin, Gracia-Abril, Portell, Teyssier, Altmann, Andrae, Bellas-Velidis,
  Benson, Berthier, Blomme, Brugaletta, Burgess, Busso, Carry, Cellino, Cheek,
  Clementini, Damerdji, Davidson, Delchambre, Dell’Oro,
  Fernández-Hernández, Galluccio, García-Lario, Garcia-Reinaldos,
  González-Núñez, Gosset, Haigron, Halbwachs, Hambly, Harrison,
  Hatzidimitriou, Heiter, Hernández, Hestroffer, Hodgkin, Holl, Janßen,
  Jevardat~de Fombelle, Jordan, Krone-Martins, Lanzafame, Löffler, Lorca,
  Manteiga, Marchal, Marrese, Moitinho, Mora, Muinonen, Osborne, Pancino,
  Pauwels, Petit, Recio-Blanco, Richards, Riello, Rimoldini, Robin, Roegiers,
  Rybizki, Sarro, Siopis, Smith, Sozzetti, Ulla, Utrilla, van Leeuwen, van
  Reeven, Abbas, Abreu~Aramburu, Accart, Aerts, Aguado, Ajaj, Altavilla,
  Álvarez, Álvarez Cid-Fuentes, Alves, Anderson, Anglada~Varela, Antoja,
  Audard, Baines, Baker, Balaguer-Núñez, Balbinot, Balog, Barache, Barbato,
  Barros, Barstow, Bartolomé, Bassilana, Bauchet, Baudesson-Stella, Becciani,
  Bellazzini, Bernet, Bertone, Bianchi, Blanco-Cuaresma, Boch, Bombrun,
  Bossini, Bouquillon, Bragaglia, Bramante, Breedt, Bressan, Brouillet,
  Bucciarelli, Burlacu, Busonero, Butkevich, Buzzi, Caffau, Cancelliere,
  Cánovas, Cantat-Gaudin, Carballo, Carlucci, Carnerero, Carrasco,
  Casamiquela, Castellani, Castro-Ginard, Castro~Sampol, Chaoul, Charlot,
  Chemin, Chiavassa, Cioni, Comoretto, Cooper, Cornez, Cowell, Crifo, Crosta,
  Crowley, Dafonte, Dapergolas, David, David, de~Laverny, De~Luise, De~March,
  De~Ridder, de~Souza, de~Teodoro, de~Torres, del Peloso, del Pozo, Delbo,
  Delgado, Delgado, Delisle, Di~Matteo, Diakite, Diener, Distefano, Dolding,
  Eappachen, Edvardsson, Enke, Esquej, Fabre, Fabrizio, Faigler, Fedorets,
  Fernique, Fienga, Figueras, Fouron, Fragkoudi, Fraile, Franke, Gai, Garabato,
  Garcia-Gutierrez, García-Torres, Garofalo, Gavras, Gerlach, Geyer, Giacobbe,
  Gilmore, Girona, Giuffrida, Gomel, Gomez, Gonzalez-Santamaria,
  González-Vidal, Granvik, Gutiérrez-Sánchez, Guy, Hauser, Haywood, Helmi,
  Hidalgo, Hilger, Hładczuk, Hobbs, Holland, Huckle, Jasniewicz, Jonker,
  Juaristi~Campillo, Julbe, Karbevska, Kervella, Khanna, Kochoska, Kontizas,
  Kordopatis, Korn, Kostrzewa-Rutkowska, Kruszyńska, Lambert, Lanza, Lasne,
  Le~Campion, Le~Fustec, Lebreton, Lebzelter, Leccia, Leclerc, Lecoeur-Taibi,
  Liao, Licata, Lindstrøm, Lister, Livanou, Lobel, Madrero~Pardo, Managau,
  Mann, Marchant, Marconi, Marcos~Santos, Marinoni, Marocco, Marshall,
  Martin~Polo, Martín-Fleitas, Masip, Massari, Mastrobuono-Battisti, Mazeh,
  McMillan, Messina, Michalik, Millar, Mints, Molina, Molinaro, Molnár,
  Montegriffo, Mor, Morbidelli, Morel, Morris, Mulone, Munoz, Muraveva, Murphy,
  Musella, Noval, Ordénovic, Orrù, Osinde, Pagani, Pagano, Palaversa,
  Palicio, Panahi, Pawlak, Peñalosa~Esteller, Penttilä, Piersimoni, Pineau,
  Plachy, Plum, Poggio, Poretti, Poujoulet, Prša, Pulone, Racero, Ragaini,
  Rainer, Raiteri, Rambaux, Ramos, Ramos-Lerate, Re~Fiorentin, Regibo, Reylé,
  Ripepi, Riva, Rixon, Robichon, Robin, Roelens, Rohrbasser, Romero-Gómez,
  Rowell, Royer, Rybicki, Sadowski, Sagristà~Sellés, Sahlmann, Salgado,
  Salguero, Samaras, Sanchez~Gimenez, Sanna, Santoveña, Sarasso, Schultheis,
  Sciacca, Segol, Segovia, Ségransan, Semeux, Shahaf, Siddiqui, Siebert,
  Siltala, Slezak, Smart, Solano, Solitro, Souami, Souchay, Spagna, Spoto,
  Steele, Steidelmüller, Stephenson, Süveges, Szabados, Szegedi-Elek, Taris,
  Tauran, Taylor, Teixeira, Thuillot, Tonello, Torra, Torra, Turon, Unger,
  Vaillant, van Dillen, Vanel, Vecchiato, Viala, Vicente, Voutsinas, Weiler,
  Wevers, Wyrzykowski, Yoldas, Yvard, Zhao, Zorec, Zucker, Zurbach, \&
  Zwitter}]{gaia_collaboration_gaia_2021}
{Gaia Collaboration}, Brown, A. G.~A., Vallenari, A., {et~al.} 2021, Astronomy
  \& Astrophysics, 649, A1, \dodoi{10.1051/0004-6361/202039657}

\bibitem[{{Gallart} {et~al.}(2019){Gallart}, {Bernard}, {Brook}, {Ruiz-Lara},
  {Cassisi}, {Hill}, \& {Monelli}}]{2019NatAs...3..932G}
{Gallart}, C., {Bernard}, E.~J., {Brook}, C.~B., {et~al.} 2019, Nature
  Astronomy, 3, 932, \dodoi{10.1038/s41550-019-0829-5}

\bibitem[{Gudin {et~al.}(2021)Gudin, Shank, Beers, Yuan, Limberg, Roederer,
  Placco, Holmbeck, Dietz, Rasmussen, Hansen, Sakari, Ezzeddine, \&
  Frebel}]{gudin_r-process_2021}
Gudin, D., Shank, D., Beers, T.~C., {et~al.} 2021, The Astrophysical Journal,
  908, 79, \dodoi{10.3847/1538-4357/abd7ed}

\bibitem[{{Han} {et~al.}(2022){Han}, {Naidu}, {Conroy}, {Bonaca}, {Zaritsky},
  {Caldwell}, {Cargile}, {Johnson}, {Chandra}, {Speagle}, {Ting}, \&
  {Woody}}]{2022arXiv220207662H}
{Han}, J.~J., {Naidu}, R.~P., {Conroy}, C., {et~al.} 2022, arXiv e-prints,
  arXiv:2202.07662.
\newblock \doarXiv{2202.07662}

\bibitem[{Hasselquist {et~al.}(2021)Hasselquist, Hayes, Lian, Weinberg,
  Zasowski, Horta, Beaton, Feuillet, Garro, Gallart, Smith, Holtzman, Minniti,
  Lacerna, Shetrone, Jönsson, Cioni, Fillingham, Cunha, O'Connell,
  Fern{\'{a}}ndez-Trincado, Mu{\~{n}}oz, Schiavon, Almeida, Anguiano, Beers,
  Bizyaev, Brownstein, Cohen, Frinchaboy, Garc{\'{\i}}a-Hern{\'{a}}ndez,
  Geisler, Lane, Majewski, Nidever, Nitschelm, Povick, Price-Whelan,
  Roman-Lopes, Rosado, Sobeck, Stringfellow, Valenzuela, Villanova, \&
  Vincenzo}]{Hasselquist_2021}
Hasselquist, S., Hayes, C.~R., Lian, J., {et~al.} 2021, The Astrophysical
  Journal, 923, 172, \dodoi{10.3847/1538-4357/ac25f9}

\bibitem[{Haywood {et~al.}(2018)Haywood, Matteo, Lehnert, Snaith, Khoperskov,
  \& G{\'{o}}mez}]{Haywood_2018}
Haywood, M., Matteo, P.~D., Lehnert, M.~D., {et~al.} 2018, The Astrophysical
  Journal, 863, 113, \dodoi{10.3847/1538-4357/aad235}

\bibitem[{Helmi(2020)}]{helmi_streams_2020}
Helmi, A. 2020, Annual Review of Astronomy and Astrophysics, 58, 205,
  \dodoi{10.1146/annurev-astro-032620-021917}

\bibitem[{Helmi {et~al.}(2018)Helmi, Babusiaux, Koppelman, Massari, Veljanoski,
  \& Brown}]{helmi_merger_2018}
Helmi, A., Babusiaux, C., Koppelman, H.~H., {et~al.} 2018, Nature, 563, 85,
  \dodoi{10.1038/s41586-018-0625-x}

\bibitem[{Helmi {et~al.}(2003)Helmi, White, \& Springel}]{Helmi_2003}
Helmi, A., White, S. D.~M., \& Springel, V. 2003, Monthly Notices of the Royal
  Astronomical Society, 339, 834, \dodoi{10.1046/j.1365-8711.2003.06227.x}

\bibitem[{{Horta} {et~al.}(2022){Horta}, {Schiavon}, {Mackereth}, {Weinberg},
  {Hasselquist}, {Feuillet}, {O'Connell}, {Anguiano}, {Allende-Prieto},
  {Beaton}, {Bizyaev}, {Cunha}, {Geisler}, {Garc{\'\i}a-Hern{\'a}ndez},
  {Holtzman}, {J{\"o}nsson}, {Lane}, {Majewski}, {M{\'e}sz{\'a}ros}, {Minniti},
  {Nitschelm}, {Shetrone}, {Smith}, \& {Zasowski}}]{2022arXiv220404233H}
{Horta}, D., {Schiavon}, R.~P., {Mackereth}, J.~T., {et~al.} 2022, arXiv
  e-prints, arXiv:2204.04233.
\newblock \doarXiv{2204.04233}

\bibitem[{Hunter(2007)}]{Hunter:2007}
Hunter, J.~D. 2007, Computing In Science \& Engineering, 9, 90

\bibitem[{Iorio \& Belokurov(2018)}]{Iorio_2018}
Iorio, G., \& Belokurov, V. 2018, Monthly Notices of the Royal Astronomical
  Society, 482, 3868, \dodoi{10.1093/mnras/sty2806}

\bibitem[{Iorio \& Belokurov(2021)}]{iorio_chemo-kinematics_2021}
---. 2021, Monthly Notices of the Royal Astronomical Society, 502, 5686,
  \dodoi{10.1093/mnras/stab005}

\bibitem[{Iorio {et~al.}(2017)Iorio, Belokurov, Erkal, Koposov, Nipoti, \&
  Fraternali}]{10.1093/mnras/stx2819}
Iorio, G., Belokurov, V., Erkal, D., {et~al.} 2017, Monthly Notices of the
  Royal Astronomical Society, 474, 2142, \dodoi{10.1093/mnras/stx2819}

\bibitem[{Jones {et~al.}(2001)Jones, Oliphant, Peterson,
  {et~al.}}]{Jones:2001ab}
Jones, E., Oliphant, T., Peterson, P., {et~al.} 2001, {SciPy}: Open source
  scientific tools for {Python}.
\newblock \url{http://www.scipy.org/}

\bibitem[{Juri{\'{c}} {et~al.}(2008)Juri{\'{c}}, Ivezi{\'{c}}, Brooks, Lupton,
  Schlegel, Finkbeiner, Padmanabhan, Bond, Sesar, Rockosi, Knapp, Gunn, Sumi,
  Schneider, Barentine, Brewington, Brinkmann, Fukugita, Harvanek, Kleinman,
  Krzesinski, Long, Eric H.~Neilsen, Nitta, Snedden, \& York}]{Juric_2008}
Juri{\'{c}}, M., Ivezi{\'{c}}, {\v{Z}}., Brooks, A., {et~al.} 2008, The
  Astrophysical Journal, 673, 864, \dodoi{10.1086/523619}

\bibitem[{Katz {et~al.}(2019)Katz, Sartoretti, Cropper, Panuzzo, Seabroke,
  Viala, Benson, Blomme, Jasniewicz, Jean-Antoine, Huckle, Smith, Baker, Crifo,
  Damerdji, David, Dolding, Fr{\'e}mat, Gosset, Guerrier, Guy, Haigron,
  Jan{\ss}en, Marchal, Plum, Soubiran, Th{\'e}venin, Ajaj, Prieto, Babusiaux,
  Boudreault, Chemin, Luche, Fabre, Gueguen, Hambly, Lasne, Meynadier, Pailler,
  Panem, Royer, Tauran, Zurbach, Zwitter, Arenou, Bossini, Gerssen, G{\'o}mez,
  Lemaitre, Leclerc, Morel, Munari, Turon, Vallenari, \& {\v
  Z}erjal}]{katz_gaia_2019}
Katz, D., Sartoretti, P., Cropper, M., {et~al.} 2019, Astronomy \&
  Astrophysics, 622, A205, \dodoi{10.1051/0004-6361/201833273}

\bibitem[{Kluyver {et~al.}(2016)Kluyver, Ragan-Kelley, P{\'e}rez, Granger,
  Bussonnier, Frederic, Kelley, Hamrick, Grout, Corlay, Ivanov, Avila, Abdalla,
  Willing, \& et~al.}]{Kluyver2016JupyterN}
Kluyver, T., Ragan-Kelley, B., P{\'e}rez, F., {et~al.} 2016, in ELPUB

\bibitem[{Koppelman {et~al.}(2020)Koppelman, Bos, \&
  Helmi}]{koppelman_massive_2020}
Koppelman, H.~H., Bos, R. O.~Y., \& Helmi, A. 2020, Astronomy \& Astrophysics,
  642, L18, \dodoi{10.1051/0004-6361/202038652}

\bibitem[{Koppelman {et~al.}(2019)Koppelman, Helmi, Massari, Price-Whelan, \&
  Starkenburg}]{koppelman_multiple_2019}
Koppelman, H.~H., Helmi, A., Massari, D., Price-Whelan, A.~M., \& Starkenburg,
  T.~K. 2019, Astronomy \& Astrophysics, 631, L9,
  \dodoi{10.1051/0004-6361/201936738}

\bibitem[{Koppelman {et~al.}(2018)Koppelman, Helmi, \&
  Veljanoski}]{koppelman_one_2018}
Koppelman, H.~H., Helmi, A., \& Veljanoski, J. 2018, The Astrophysical Journal,
  860, L11, \dodoi{10.3847/2041-8213/aac882}

\bibitem[{{Kruijssen} {et~al.}(2020){Kruijssen}, {Pfeffer}, {Chevance},
  {Bonaca}, {Trujillo-Gomez}, {Bastian}, {Reina-Campos}, {Crain}, \&
  {Hughes}}]{2020MNRAS.498.2472K}
{Kruijssen}, J.~M.~D., {Pfeffer}, J.~L., {Chevance}, M., {et~al.} 2020, \mnras,
  498, 2472, \dodoi{10.1093/mnras/staa2452}

\bibitem[{Lancaster {et~al.}(2019)Lancaster, Koposov, Belokurov, Evans, \&
  Deason}]{lancaster_halos_2019}
Lancaster, L., Koposov, S.~E., Belokurov, V., Evans, N.~W., \& Deason, A.~J.
  2019, Monthly Notices of the Royal Astronomical Society, 486, 378,
  \dodoi{10.1093/mnras/stz853}

\bibitem[{Limberg {et~al.}(2021)Limberg, Santucci, Rossi, Shank, Placco, Beers,
  Schlaufman, Casey, Perottoni, \& Lee}]{limberg_targeting_2021}
Limberg, G., Santucci, R.~M., Rossi, S., {et~al.} 2021, The Astrophysical
  Journal, 913, 11, \dodoi{10.3847/1538-4357/abeefe}

\bibitem[{{Lindegren} {et~al.}(2021){Lindegren}, {Bastian}, {Biermann},
  {Bombrun}, {de Torres}, {Gerlach}, {Geyer}, {Hern{\'a}ndez}, {Hilger},
  {Hobbs}, {Klioner}, {Lammers}, {McMillan}, {Ramos-Lerate},
  {Steidelm{\"u}ller}, {Stephenson}, \& {van Leeuwen}}]{2021A&A...649A...4L}
{Lindegren}, L., {Bastian}, U., {Biermann}, M., {et~al.} 2021, \aap, 649, A4,
  \dodoi{10.1051/0004-6361/202039653}

\bibitem[{M.~Price-Whelan(2017)}]{m_price-whelan_gala_2017}
M.~Price-Whelan, A. 2017, The Journal of Open Source Software, 2, 388,
  \dodoi{10.21105/joss.00388}

\bibitem[{Mackereth \& Bovy(2020)}]{10.1093/mnras/staa047}
Mackereth, J.~T., \& Bovy, J. 2020, Monthly Notices of the Royal Astronomical
  Society, 492, 3631, \dodoi{10.1093/mnras/staa047}

\bibitem[{Mackereth {et~al.}(2019)Mackereth, Schiavon, Pfeffer, Hayes, Bovy,
  Anguiano, Allende~Prieto, Hasselquist, Holtzman, Johnson, Majewski,
  O'Connell, Shetrone, Tissera, \& Fernández-Trincado}]{mackereth_origin_2019}
Mackereth, J.~T., Schiavon, R.~P., Pfeffer, J., {et~al.} 2019, Monthly Notices
  of the Royal Astronomical Society, 482, 3426, \dodoi{10.1093/mnras/sty2955}

\bibitem[{{Majewski} {et~al.}(2017){Majewski}, {Schiavon}, {Frinchaboy},
  {Allende Prieto}, {Barkhouser}, {Bizyaev}, {Blank}, {Brunner}, {Burton},
  {Carrera}, {Chojnowski}, {Cunha}, {Epstein}, {Fitzgerald}, {Garc{\'\i}a
  P{\'e}rez}, {Hearty}, {Henderson}, {Holtzman}, {Johnson}, {Lam}, {Lawler},
  {Maseman}, {M{\'e}sz{\'a}ros}, {Nelson}, {Nguyen}, {Nidever}, {Pinsonneault},
  {Shetrone}, {Smee}, {Smith}, {Stolberg}, {Skrutskie}, {Walker}, {Wilson},
  {Zasowski}, {Anders}, {Basu}, {Beland}, {Blanton}, {Bovy}, {Brownstein},
  {Carlberg}, {Chaplin}, {Chiappini}, {Eisenstein}, {Elsworth}, {Feuillet},
  {Fleming}, {Galbraith-Frew}, {Garc{\'\i}a}, {Garc{\'\i}a-Hern{\'a}ndez},
  {Gillespie}, {Girardi}, {Gunn}, {Hasselquist}, {Hayden}, {Hekker}, {Ivans},
  {Kinemuchi}, {Klaene}, {Mahadevan}, {Mathur}, {Mosser}, {Muna}, {Munn},
  {Nichol}, {O'Connell}, {Parejko}, {Robin}, {Rocha-Pinto}, {Schultheis},
  {Serenelli}, {Shane}, {Silva Aguirre}, {Sobeck}, {Thompson}, {Troup},
  {Weinberg}, \& {Zamora}}]{2017AJ....154...94M}
{Majewski}, S.~R., {Schiavon}, R.~P., {Frinchaboy}, P.~M., {et~al.} 2017, \aj,
  154, 94, \dodoi{10.3847/1538-3881/aa784d}

\bibitem[{{Massari} {et~al.}(2019){Massari}, {Koppelman}, \&
  {Helmi}}]{2019A&A...630L...4M}
{Massari}, D., {Koppelman}, H.~H., \& {Helmi}, A. 2019, \aap, 630, L4,
  \dodoi{10.1051/0004-6361/201936135}

\bibitem[{{McWilliam}(1997)}]{1997ARA&A..35..503M}
{McWilliam}, A. 1997, \araa, 35, 503, \dodoi{10.1146/annurev.astro.35.1.503}

\bibitem[{{Montalb{\'a}n} {et~al.}(2021){Montalb{\'a}n}, {Mackereth}, {Miglio},
  {Vincenzo}, {Chiappini}, {Buldgen}, {Mosser}, {Noels}, {Scuflaire}, {Vrard},
  {Willett}, {Davies}, {Hall}, {Nielsen}, {Khan}, {Rendle}, {van Rossem},
  {Ferguson}, \& {Chaplin}}]{2021NatAs...5..640M}
{Montalb{\'a}n}, J., {Mackereth}, J.~T., {Miglio}, A., {et~al.} 2021, Nature
  Astronomy, 5, 640, \dodoi{10.1038/s41550-021-01347-7}

\bibitem[{Myeong {et~al.}(2018)Myeong, Evans, Belokurov, Sanders, \&
  Koposov}]{Myeong_2018}
Myeong, G.~C., Evans, N.~W., Belokurov, V., Sanders, J.~L., \& Koposov, S.~E.
  2018, The Astrophysical Journal, 863, L28, \dodoi{10.3847/2041-8213/aad7f7}

\bibitem[{{Myeong} {et~al.}(2019){Myeong}, {Vasiliev}, {Iorio}, {Evans}, \&
  {Belokurov}}]{2019MNRAS.488.1235M}
{Myeong}, G.~C., {Vasiliev}, E., {Iorio}, G., {Evans}, N.~W., \& {Belokurov},
  V. 2019, \mnras, 488, 1235, \dodoi{10.1093/mnras/stz1770}

\bibitem[{Naidu {et~al.}(2020)Naidu, Conroy, Bonaca, Johnson, Ting, Caldwell,
  Zaritsky, \& Cargile}]{naidu_evidence_2020}
Naidu, R.~P., Conroy, C., Bonaca, A., {et~al.} 2020, The Astrophysical Journal,
  901, 48, \dodoi{10.3847/1538-4357/abaef4}

\bibitem[{{Naidu} {et~al.}(2021){Naidu}, {Conroy}, {Bonaca}, {Zaritsky},
  {Weinberger}, {Ting}, {Caldwell}, {Tacchella}, {Han}, {Speagle}, \&
  {Cargile}}]{naidu_reconstructing_2021}
{Naidu}, R.~P., {Conroy}, C., {Bonaca}, A., {et~al.} 2021, \apj, 923, 92,
  \dodoi{10.3847/1538-4357/ac2d2d}

\bibitem[{{Necib} {et~al.}(2019){Necib}, {Lisanti}, \&
  {Belokurov}}]{Necib:2018iwb}
{Necib}, L., {Lisanti}, M., \& {Belokurov}, V. 2019, \apj, 874, 3,
  \dodoi{10.3847/1538-4357/ab095b}

\bibitem[{Necib {et~al.}(2019)Necib, Lisanti, Garrison-Kimmel, Wetzel,
  Sanderson, Hopkins, Faucher-Giguère, \& Kereš}]{necib_under_2019}
Necib, L., Lisanti, M., Garrison-Kimmel, S., {et~al.} 2019, The Astrophysical
  Journal, 883, 27, \dodoi{10.3847/1538-4357/ab3afc}

\bibitem[{Newberg {et~al.}(2002)Newberg, Yanny, Rockosi, Grebel, Rix,
  Brinkmann, Csabai, Hennessy, Hindsley, Ibata, Ivezi{\'{c}}, Lamb, Nash,
  Odenkirchen, Rave, Schneider, Smith, Stolte, \& York}]{Newberg_2002}
Newberg, H.~J., Yanny, B., Rockosi, C., {et~al.} 2002, The Astrophysical
  Journal, 569, 245, \dodoi{10.1086/338983}

\bibitem[{{Perez} \& {Granger}(2007)}]{PER-GRA:2007}
{Perez}, F., \& {Granger}, B.~E. 2007, Computing in Science and Engineering, 9,
  21, \dodoi{10.1109/MCSE.2007.53}

\bibitem[{Queiroz {et~al.}(2020)Queiroz, Anders, Chiappini, Khalatyan,
  Santiago, Steinmetz, Valentini, Miglio, Bossini, Barbuy, Minchev, Minniti,
  García~Hernández, Schultheis, Beaton, Beers, Bizyaev, Brownstein, Cunha,
  Fernández-Trincado, Frinchaboy, Lane, Majewski, Nataf, Nitschelm, Pan,
  Roman-Lopes, Sobeck, Stringfellow, \& Zamora}]{queiroz_bulge_2020}
Queiroz, A. B.~A., Anders, F., Chiappini, C., {et~al.} 2020, Astronomy \&
  Astrophysics, 638, A76, \dodoi{10.1051/0004-6361/201937364}

\bibitem[{{Rix} \& {Bovy}(2013)}]{2013A&ARv..21...61R}
{Rix}, H.-W., \& {Bovy}, J. 2013, \aapr, 21, 61,
  \dodoi{10.1007/s00159-013-0061-8}

\bibitem[{Robertson {et~al.}(2005)Robertson, Bullock, Font, Johnston, \&
  Hernquist}]{Robertson_2005}
Robertson, B., Bullock, J.~S., Font, A.~S., Johnston, K.~V., \& Hernquist, L.
  2005, The Astrophysical Journal, 632, 872, \dodoi{10.1086/452619}

\bibitem[{Robitaille {et~al.}(2013)}]{Robitaille:2013mpa}
Robitaille, T.~P., {et~al.} 2013, Astron. Astrophys., 558, A33,
  \dodoi{10.1051/0004-6361/201322068}

\bibitem[{{Simion} {et~al.}(2014){Simion}, {Belokurov}, {Irwin}, \&
  {Koposov}}]{2014MNRAS.440..161S}
{Simion}, I.~T., {Belokurov}, V., {Irwin}, M., \& {Koposov}, S.~E. 2014,
  \mnras, 440, 161, \dodoi{10.1093/mnras/stu133}

\bibitem[{{Simion} {et~al.}(2019){Simion}, {Belokurov}, \&
  {Koposov}}]{2019MNRAS.482..921S}
{Simion}, I.~T., {Belokurov}, V., \& {Koposov}, S.~E. 2019, \mnras, 482, 921,
  \dodoi{10.1093/mnras/sty2744}

\bibitem[{{Tolstoy} {et~al.}(2009){Tolstoy}, {Hill}, \&
  {Tosi}}]{2009ARA&A..47..371T}
{Tolstoy}, E., {Hill}, V., \& {Tosi}, M. 2009, \araa, 47, 371,
  \dodoi{10.1146/annurev-astro-082708-101650}

\bibitem[{{van der Walt} {et~al.}(2011){van der Walt}, {Colbert}, \&
  {Varoquaux}}]{numpy:2011}
{van der Walt}, S., {Colbert}, S.~C., \& {Varoquaux}, G. 2011, Computing in
  Science and Engineering, 13, 22, \dodoi{10.1109/MCSE.2011.37}

\bibitem[{Vincenzo {et~al.}(2019)Vincenzo, Spitoni, Calura, Matteucci,
  Silva Aguirre, Miglio, \& Cescutti}]{10.1093/mnrasl/slz070}
Vincenzo, F., Spitoni, E., Calura, F., {et~al.} 2019, Monthly Notices of the
  Royal Astronomical Society: Letters, 487, L47, \dodoi{10.1093/mnrasl/slz070}

\bibitem[{Vivas {et~al.}(2001)Vivas, Zinn, Andrews, Bailyn, Baltay, Coppi,
  Ellman, Girard, Rabinowitz, Schaefer, Shin, Snyder, Sofia, van Altena, Abad,
  Bongiovanni, Brice{\~{n}}o, Bruzual, Prugna, Herrera, Magris, Mateu, Pacheco,
  S{\'{a}}nchez, S{\'{a}}nchez, Schenner, Stock, Vicente, Vieira,
  Ferr{\'{\i}}n, Hernandez, Gebhard, Honeycutt, Mufson, Musser, \&
  Rengstorf}]{Vivas_2001}
Vivas, A.~K., Zinn, R., Andrews, P., {et~al.} 2001, The Astrophysical Journal,
  554, L33, \dodoi{10.1086/320915}

\bibitem[{{W}es {M}c{K}inney(2010)}]{mckinney-proc-scipy-2010}
{W}es {M}c{K}inney. 2010, in {P}roceedings of the 9th {P}ython in {S}cience
  {C}onference, ed. {S}t\'efan van~der {W}alt \& {J}arrod {M}illman, 56 -- 61,
  \dodoi{10.25080/Majora-92bf1922-00a}

\bibitem[{{Wheeler} {et~al.}(1989){Wheeler}, {Sneden}, \&
  {Truran}}]{1989ARA&A..27..279W}
{Wheeler}, J.~C., {Sneden}, C., \& {Truran}, James~W., J. 1989, \araa, 27, 279,
  \dodoi{10.1146/annurev.aa.27.090189.001431}

\bibitem[{White \& Frenk(1991)}]{white_galaxy_1991}
White, S. D.~M., \& Frenk, C.~S. 1991, The Astrophysical Journal, 379, 52,
  \dodoi{10.1086/170483}

\bibitem[{Yuan {et~al.}(2020)Yuan, Chang, Beers, \& Huang}]{yuan_low-mass_2020}
Yuan, Z., Chang, J., Beers, T.~C., \& Huang, Y. 2020, The Astrophysical
  Journal, 898, L37, \dodoi{10.3847/2041-8213/aba49f}

\end{thebibliography}
\bibliographystyle{aasjournal}

\end{document}